\documentclass[prd,aps,twocolumn,bibnotes,nofootinbib,preprintnumbers,floatfix]{revtex4-1}
\pdfoutput=1
 
\usepackage{amsmath}
\usepackage{bbold}
\usepackage[normalem]{ulem}

\usepackage{graphicx}
\usepackage[small]{subfigure}
\usepackage{hyperref}
\usepackage{xspace}
\usepackage{braket}
\usepackage{placeins}
\usepackage{xcolor}

\DeclareRobustCommand{\eq}[1]{Eq.~\eqref{eq:#1}}
\DeclareRobustCommand{\eqs}[2]{Eqs.~\eqref{eq:#1} and \eqref{eq:#2}}
\DeclareRobustCommand{\fig}[1]{Fig.~\ref{fig:#1}}

\DeclareRobustCommand{\sec}[1]{Sec.~\ref{sec:#1}}

\DeclareRobustCommand{\app}[1]{appendix~\ref{app:#1}}
\DeclareRobustCommand{\mycite}[1]{Ref.~\cite{#1}}
\DeclareRobustCommand{\mycites}[1]{Refs.~\cite{#1}}

\newcommand{\df}{\mathrm{d}}
\newcommand{\img}{\mathrm{i}}
\newcommand{\eps}{\epsilon}
\newcommand{\cA}{\mathcal{A}}
\newcommand{\cO}{\mathcal{O}}
\newcommand{\GeV}{\,\mathrm{GeV}}
\newcommand{\nn}{\nonumber}
\newcommand{\LQCD}{\Lambda_{\rm QCD}}
\newcommand{\Ecm}{E_{\rm cm}}
\newcommand{\bn}{{\bar{n}}}
\newcommand{\as}{\alpha_s}
\newcommand{\bt}{\vec b_T}

\newcommand{\qt}{\vec q_T}
\newcommand{\GammaC}{\Gamma_{\rm cusp}}
\newcommand{\MS}{\overline{\mathrm{MS}}}
\newcommand{\ns}{{\text{ns}}}
\newcommand\TMD{\mathrm{TMD}}

\allowdisplaybreaks

\begin{document}

\preprint{MIT--CTP 5049}

\title{Determining the Nonperturbative Collins-Soper Kernel From Lattice QCD}

\author{Markus A.~Ebert}
\email{ebert@mit.edu}
\author{Iain W.~Stewart}
\email{iains@mit.edu}
\author{Yong Zhao}
\email{yzhaoqcd@mit.edu}
\affiliation{\vspace{0.2cm} Center for Theoretical Physics, Massachusetts Institute of Technology, Cambridge, Massachusetts 02139, USA}

\begin{abstract}

At small transverse momentum $q_T$, transverse-momentum dependent parton distribution functions (TMDPDFs) arise as 
genuinely nonperturbative 
objects that describe Drell-Yan like processes in hadron collisions as well as semi-inclusive deep-inelastic scattering.
TMDPDFs naturally depend on the hadron momentum, and the associated evolution is determined by the Collins-Soper equation.
For $q_T \sim \LQCD$ the corresponding evolution kernel (or anomalous dimension) is nonperturbative and must be determined as an independent ingredient in order to relate TMDPDFs at different  scales.
We propose a method to extract this kernel using lattice QCD and the Large-Momentum Effective Theory, where the physical TMD correlation involving light-like paths is approximated by a quasi-TMDPDF, defined using equal-time correlation functions with a large-momentum hadron state.
The kernel is determined from a ratio of quasi-TMDPDFs extracted at different hadron momenta.

\end{abstract}

\maketitle

\section{Introduction}

In the past decades, advances in theory and experiment have made it possible to explore the structure of the proton beyond the simplest longitudinal momentum distributions. Key observables are transverse momentum distributions (TMDs), which measure the intrinsic transverse momentum $q_T$ of partons in the proton, as well as describing the probability to produce particles at larger $q_T$ in high energy collisions. These TMDs are probed directly by experiments on Drell-Yan, semi-inclusive deep inelastic scattering (SIDIS), and other processes.
Recently, progress has been made in determining TMDPDFs by using lattice QCD \cite{Hagler:2009mb,Musch:2010ka,Musch:2011er,Engelhardt:2015xja,Yoon:2016dyh,Yoon:2017qzo} to study equal-time correlators.
Such correlators are a key ingredient in the large-momentum effective theory (LaMET)~\cite{Ji:2013dva,Ji:2014gla}, where one computes a lightcone correlator using an equal-time correlator in a boosted proton state.  For TMDPDFs the first theoretical studies in LaMET have been carried out in Refs.~\cite{Ji:2014hxa,Ji:2018hvs,Ebert:2019okf}.

The TMDPDF $f_i^\TMD(x, \bt, \mu, \zeta)$ for a parton of flavor $i$  depends on $x$, the fraction of the hadron momentum carried by the struck parton, $\bt$, the Fourier-conjugate of the transverse momentum $\qt$, and on a virtuality scale $\mu$. In addition, it depends on a scale $\zeta$, which is related to the momentum of the hadron or equivalently the hard scale of the scattering process.
Measuring nonperturbative TMDPDFs, whether from experiment or lattice, thus requires to specify the scales $(\mu_0,\zeta_0)$ where the TMDPDF is extracted.
For instance, lattice calculations are restricted to $\mu_0^2 \sim \zeta_0 \sim \cO(4\:\GeV^2)$ due to finite lattice spacing,
while for example the application to Drell-Yan production uses $\mu^2 \sim \zeta \sim m_Z^2 \approx (91~\GeV)^2$.
The TMDPDFs thus need be evolved from $(\mu_0,\zeta_0)$ to the phenomenologically relevant scales $(\mu,\zeta)$,
using 
\begin{align} \label{eq:evolution}
 &f_i^\TMD(x, \bt, \mu, \zeta) = f_i^\TMD(x, \bt, \mu_0, \zeta_0)
 \nn\\*&\quad\times
 \exp\biggl[ \int_{\mu_0}^\mu \frac{\df\mu'}{\mu'} \gamma_\mu^i(\mu',\zeta_0) \biggr]
 \exp\biggl[ \frac12 \gamma_\zeta^i(\mu,b_T) \ln\frac{\zeta}{\zeta_0} \biggr]
\,.\end{align}
The first exponential in \eq{evolution} is the $\mu$ evolution and the second exponential is the Collins-Soper evolution in $\zeta$ \cite{Collins:1981uk,Collins:1981va,Collins:1984kg} in the formulation of~\mycite{Collins:1350496}, with $\gamma_\mu^i$ and $\gamma_\zeta^i$ being the associated anomalous dimensions. Here $\gamma_\zeta^i$ is the Collins-Soper kernel, often denoted by $\tilde K$.

The $\mu$ evolution in \eq{evolution} is perturbative as long as both $\mu_0, \mu \gg \LQCD$, analogous to the perturbative DGLAP evolution for collinear PDFs.
In contrast, the Collins-Soper kernel involves the $b_T$-dependent anomalous dimension $\gamma_\zeta^i(\mu,b_T)$,
which becomes nonperturbative in the region $b_T^{-1} \sim q_T \sim \LQCD$, even if $\mu \gg \LQCD$. Relating the nonperturbative TMDPDF extracted at some reference scales $(\mu_0,\zeta_0)$ to the phenomenologically relevant scales $(\mu,\zeta)$ thus crucially relies on the nonperturbative knowledge of the Collins-Soper kernel.

Due to the simple form of \eq{evolution}, the Collins-Soper evolution can be factored out by taking the ratio of two TMDPDFs extracted at different values $\zeta_1 \ne \zeta_2$,
\begin{align} \label{eq:gamma_zeta_measurement}
 \gamma_\zeta^i(\mu,b_T)
 = \frac{2}{\ln(\zeta_1/\zeta_2)} \,
  \ln\frac{f^\TMD_i(x, \bt, \mu, \zeta_1)}{f^\TMD_i(x, \bt, \mu, \zeta_2)}
\,.\end{align}
One option is to extract $\gamma_\zeta^{i}(\mu,b_T)$ experimentally. For example, one can use Drell-Yan production at small $q_T\ll Q$, where $Q$ is the invariant mass of the Drell-Yan pair, and use different values of $Q$ to obtain $f^\TMD_{i=q}(x, \bt, \mu, \zeta)$ at different $\zeta$ values. This can be done using results from global fits, see e.g.\ \mycites{Bacchetta:2017gcc,Scimemi:2017etj}. Experimentally, the TMDPDF is extracted as a function of $\vec q_T$, which makes it challenging to use \eq{gamma_zeta_measurement} since it requires the Fourier transformation into $\vec b_T$ space. 
Interestingly, \eq{gamma_zeta_measurement} is independent of the momentum fraction $x$ and choice of $\zeta_{1,2}$,
which is useful to assess associated systematics and to validate the applicability of TMD factorization.

In this paper, we propose a first-principle method of determining the nonperturbative $\gamma_\zeta^q$ using lattice QCD.
A potential benefit is that one has, in principle, more control over the systematics in the calculation.
The TMDPDFs in \eq{gamma_zeta_measurement} are not directly computable on the lattice because they involve time dependent operators with Wilson lines on (or close to) the light cone. We therefore consider the LaMET approach for calculating the ratio in \eq{gamma_zeta_measurement} from lattice QCD, which will involve additional perturbative matching corrections. Our proposed formula for  $\gamma_\zeta^q(\mu,b_T)$ is analogous to \eq{gamma_zeta_measurement} and is given in \sec{gamma_zeta}.

Below in \sec{tmd_review}, we briefly review the definition of TMDPDFs in the context of TMD factorization for proton-proton collisions.
In \sec{quasi_tmd}, we present the construction of quasi-TMDPDFs using LaMET that are computable on lattice, and briefly address issues arising from the presence of both collinear and soft matrix elements.
The strategy to extract $\gamma_\zeta^q$ from lattice is presented in \sec{gamma_zeta}.
We conclude in \sec{conclusions}.

\section{TMD Factorization}
\label{sec:tmd_review}

TMD factorization was originally derived by Collins, Soper and Sterman (CSS)
in \mycites{Collins:1981uk,Collins:1981va,Collins:1984kg} and extended by Collins in \mycite{Collins:1350496}.
The cancellation of potentially factorization-violating Glauber contributions has been shown in
\mycites{Collins:1985ue,Collins:1988ig,Collins:1989gx,Collins:1350496,Diehl:2015bca}.
The factorization has also been considered in the framework of Soft-Collinear Effective Theory (SCET)
\cite{Bauer:2000ew, Bauer:2000yr, Bauer:2001ct, Bauer:2001yt}
by various authors \cite{Becher:2010tm, Becher:2011xn, Becher:2012yn, GarciaEchevarria:2011rb, Echevarria:2012js, Echevarria:2014rua, Chiu:2012ir},
see e.g.\ \mycite{Collins:2017oxh} for a detailed discussion of the different approaches.

We consider the production of a color-singlet final state $F$
with invariant mass $Q$, rapidity $Y$ and small transverse momentum $q_T=|\qt| \ll Q$
in the scattering of two energetic protons moving close to the
$n^\mu = (1,0,0,1)$ and $\bn^\mu = (1,0,0,-1)$ directions with a center of mass energy $\Ecm$.
In the limit $q_T \ll Q$ the cross section can be factorized as
\begin{align} \label{eq:sigma}
 &\frac{\df\sigma}{\df Q \df Y \df^2\qt}
 = \sum_{i,j} H_{ij}(Q,\mu) \int\!\! \df^2\bt \, e^{\img \bt \cdot \qt}
 \nn\\* &\times
 f^\TMD_i(x_a, \bt, \mu, \zeta_a) \,
 f^\TMD_j(x_b, \bt, \mu, \zeta_b)
\,,\end{align}
which holds up to power corrections of relative order $q_T^2/Q^2$ and $\LQCD^2/Q^2$, but remains valid in the nonperturbative regime $q_T\sim \LQCD$. 
Here, $H_{ij}$ is the hard function describing virtual corrections to the underlying hard process $ij\to F$, where $i,j$ are the parton flavors
(for gluon-induced process, $ij=gg$, the hard function and the TMDPDFs carry helicity indices, which are suppressed here),
$x_{a,b} = Q e^{\pm Y}/\Ecm$ are the longitudinal momentum fractions carried by the struck partons,
and $f^\TMD$ are the TMDPDFs in Fourier space.
(Our $H$ and $f^\TMD$s agree with the definitions used in \mycites{Collins:1350496,Catani:2000vq,Becher:2010tm,Becher:2011dz,Chiu:2012ir,GarciaEchevarria:2011rb,Gehrmann:2012ze,Li:2016axz,Echevarria:2016scs}, but differ from those of \mycites{Collins:1981uk,Collins:1981va,Collins:1984kg,Ji:2004wu}, see \mycites{Prokudin:2015ysa,Collins:2017oxh} for relations.) Note that these physical TMDs involve staple Wilson line paths, and hence differ from the straight line paths considered in \mycites{Hagler:2009mb,Musch:2010ka,Radyushkin:2017cyf}.
A similar result is obtained for SIDIS, $e p \to h + X$, where the transverse momentum of the  hadron $h$ is measured  by a TMD fragmentation function rather than a TMDPDF,
which obeys the same Collins-Soper equation \cite{Collins:1350496}.

An important feature of TMDs is that bare matrix elements not only suffer from UV divergences regulated by $\epsilon$ (for example in dimensional regularization with $d=4-2\eps$), but also from \emph{rapidity divergences} (also known as lightcone divergences) which require another regulator \cite{Collins:1981uk,Collins:1992tv,Collins:2008ht,Becher:2010tm,GarciaEchevarria:2011rb,Chiu:2011qc,Chiu:2012ir} that we will generically denote by $\tau$.
Many such regulators have been suggested in the literature \cite{Collins:1981uk,Collins:1350496,Ji:2004wu,Beneke:2003pa, Chiu:2007yn, Becher:2011dz,Chiu:2011qc, Chiu:2012ir,Chiu:2009yx, GarciaEchevarria:2011rb,Li:2016axz}. As a consequence, TMDPDFs depend on both the renormalization scale $\mu$ and the parameter $\zeta$.  In \eq{sigma} $\mu$ formally cancels between the $H$ and $f^\TMD$ functions, but in practice is chosen as $\mu \sim Q$ to avoid large logarithms $\ln(Q/\mu)$ in the hard function.
The values for $\zeta_a$ and $\zeta_b$ are not fixed individually, but their product is fixed to $\zeta_a \zeta_b = Q^4$.
This forces one to evaluate the TMDPDFs at $\mu^2 \sim \zeta \sim Q^2$, and requires the use of \eq{evolution} to evolve them from some other scales like $(\mu_0^2, \zeta_0) \sim \cO(4\: \GeV^2)$ where they are nonperturbatively determined (or parameterized).

A definition of the quark TMDPDF consistent with Refs.~\cite{Collins:1350496,Chiu:2012ir,GarciaEchevarria:2011rb} can be given by
\begin{align}
 \label{eq:tmdpdf}
 f^\TMD_q(x, \bt, \mu , \zeta) &=  \lim_{\substack{\epsilon\to 0 \\ \tau\to 0} }\,
 Z_{\rm uv}(\mu,\zeta,\eps)\,
  B_{q}(x, \bt, \eps, \tau, \zeta)
  \nn\\
  &\qquad\quad \times \Delta_S^q(b_T,\eps,\tau) 
\,.\end{align}
Here $Z_{\rm uv}$ is the ultraviolet (UV) renormalization factor, and we refer to $B_{q}$ and $\Delta^q_{S}$ as the bare beam function (where we follow the naming scheme of~\mycite{Stewart:2009yx}) and soft factor, respectively, to distinguish them from the TMDPDF $f_q^\TMD$.  \mycites{Collins:1350496,Chiu:2012ir,GarciaEchevarria:2011rb} use different definitions for $\tau$, $B_{q}$, and $\Delta_S^q$ but all choices yield the same $f_q^\TMD$. As $\tau\to 0$ only the combination $1/\tau - \ln\sqrt{\zeta}$ shows up in the bare function $B_{q}$. Importantly, $1/\tau$ divergences cancel out in \eq{tmdpdf} yielding a well-defined TMDPDF. A remnant of this cancellation is the appearance of $\zeta$ in $f_{q}^\TMD$. In \eq{sigma} we have
\begin{align} \label{eq:zeta_Collins}
\zeta_a = x_a^2 (P_a^-)^2 e^{-2y_n} \,,\quad \zeta_b = x_b^2 (P_b^+)^2 e^{2y_n}
\,,\end{align}
where $P_{a,b}$ are the hadron momenta and $y_n$ is an arbitrary parameter controlling the split of soft radiation into the TMDPDFs (where the precise specification of $y_n$ differs in \mycites{Collins:1350496,Chiu:2012ir,GarciaEchevarria:2011rb}).
Their product is always fixed to
\begin{equation} \label{eq:zeta_product}
\zeta_a \zeta_b = Q^4 = (x_a x_b \Ecm^2)^2
\,.
\end{equation}
\eqs{zeta_Collins}{zeta_product} involve the momentum fractions $x_a$ and $x_b$,
and thus are specified in momentum space.

A well-known example of \eq{tmdpdf} is Collins' regulator~\cite{Collins:1350496} where Wilson lines are taken off the lightcone, and the soft factor is defined as
\begin{align}
 \Delta_S^q(b_T,\epsilon,\tau) &
  = \lim_{y_A\to \infty} \sqrt{\frac{S^{y_A,y_n}}{S^{y_A,y_B}S^{y_n,y_B}}}
 \nn\\&
  = \frac{1}{\sqrt{S^q_{\rm C}\bigl(b_T,\epsilon,2y_n-2y_B\bigr)}}
\,,\end{align}
(see \mycite{Buffing:2017mqm} for the last equality).
Here, $S^{y_a,y_b}$ denotes a soft matrix element with Wilson lines parameterized by the rapidities $y_a$ and $y_b$,
and $y_A$ and $y_B$ are the rapidities of the Wilson lines entering $B_q$ for the $n$-collinear proton and $\bn$-collinear proton, respectively.
The regulator is given by $\tau =1/(y_B-y_n)$ with $y_B\to - \infty$.
For Collins' beam function we have
\begin{align}
B_q(x,\bt,\epsilon,\tau,\zeta) &= B_q^{\rm C}(x,\bt,\epsilon,y_P-y_B) \,,
\end{align}
which only depends on the rapidity difference $y_P-y_B$ since $1/\tau - \ln\sqrt{\zeta} = y_B-y_P -\ln (x\,m_P)$, where $m_P$ and $y_P$ are the proton mass and rapidity, and $P^-=m_P\, e^{y_P}$ is the proton momentum.

In the EIS scheme~\cite{GarciaEchevarria:2011rb,Echevarria:2012js} one regulates eikonal propagators by basically shifting $1/(k^\pm + \img0) \to 1/(k^\pm + \img \delta^\pm)$. In this scheme there is a soft function $S$ and two zero-bin subtraction~\cite{Manohar:2006nz} functions $S_0$ which avoid double counting between the soft function $S$ and the beam function $B$. The $S$ and $S_0$ appear together as a multiplicative factor of $S/(S_0 S_0)=1/S=1/(\sqrt{S}\sqrt{S})$. With the $\delta^\pm$ regulators one therefore defines the soft factor appearing in the TMD as
\begin{align}
 \Delta_S^q(b_T,\eps,\tau) = \frac{1}{\sqrt{S^q_{\rm EIS}\bigl(b_T, \eps, \delta^- e^{-y_n}\bigr)}}
\,.\end{align}
Here the regulator is $1/\tau = \ln(\delta^- e^{-y_n})$. In this scheme 
\begin{align}
  B_q(x,\bt,\epsilon,\tau,\zeta) &= B_q^{\rm EIS} \bigl(x,\bt,\epsilon, \delta^-/( x P^-) \bigr) \,,
\end{align}
where $B_q^{\rm EIS} \equiv J_n$ in the notation of \mycite{Echevarria:2012js} and $1/\tau -\ln\sqrt{\zeta} = \ln[ \delta^-/(x P^-)]$.

Finally, in the $\eta$-regulator and scheme of CJNR in \mycite{Chiu:2012ir} the Wilson lines in the proton and vacuum matrix elements are regulated with factors of $|2k^z|^{-\eta}$,
where $k$ is the momentum of gluons emitted from the Wilson line,
and we have $\tau = \eta$. In this scheme the zero-bin subtraction functions vanish, so $S/(S_0 S_0) = S = \sqrt{S}\sqrt{S}$. Therefore the soft factor is
\begin{align}
\Delta_S^q(b_T,\epsilon,\tau) = \sqrt{ S^q_{\rm CJNR}(b_T,\epsilon,\eta)} 
 \,.
\end{align}
In this approach the regulator is chosen to act symmetrically on the two proton matrix elements, so $y_n=0$,  $\zeta = (x P^-)^2$, and the bare beam function is
\begin{align}
  B_q(x,\bt,\epsilon,\tau,\zeta) 
    &= B_q^{\rm CJNR}\bigl(x,\bt,\epsilon,\eta, (x P^-)^2\bigr)
  \,.
\end{align}

The exponential regulator defined in \mycite{Li:2016axz}, and used for the three-loop perturbative computation of $\gamma_\zeta$ in \mycite{Li:2016ctv}, has a similar functional dependence in $S$ and $B_q$ as the $\eta$-regulator  (though the zero-bin subtraction functions $S_0$ do not vanish in this case). So far this scheme has only been used for perturbative calculations, but it also belongs in the set of TMDPDF definitions governed by \eq{tmdpdf}.

\begin{figure*}[t]
	\centering
	\includegraphics[width=0.35\textwidth]{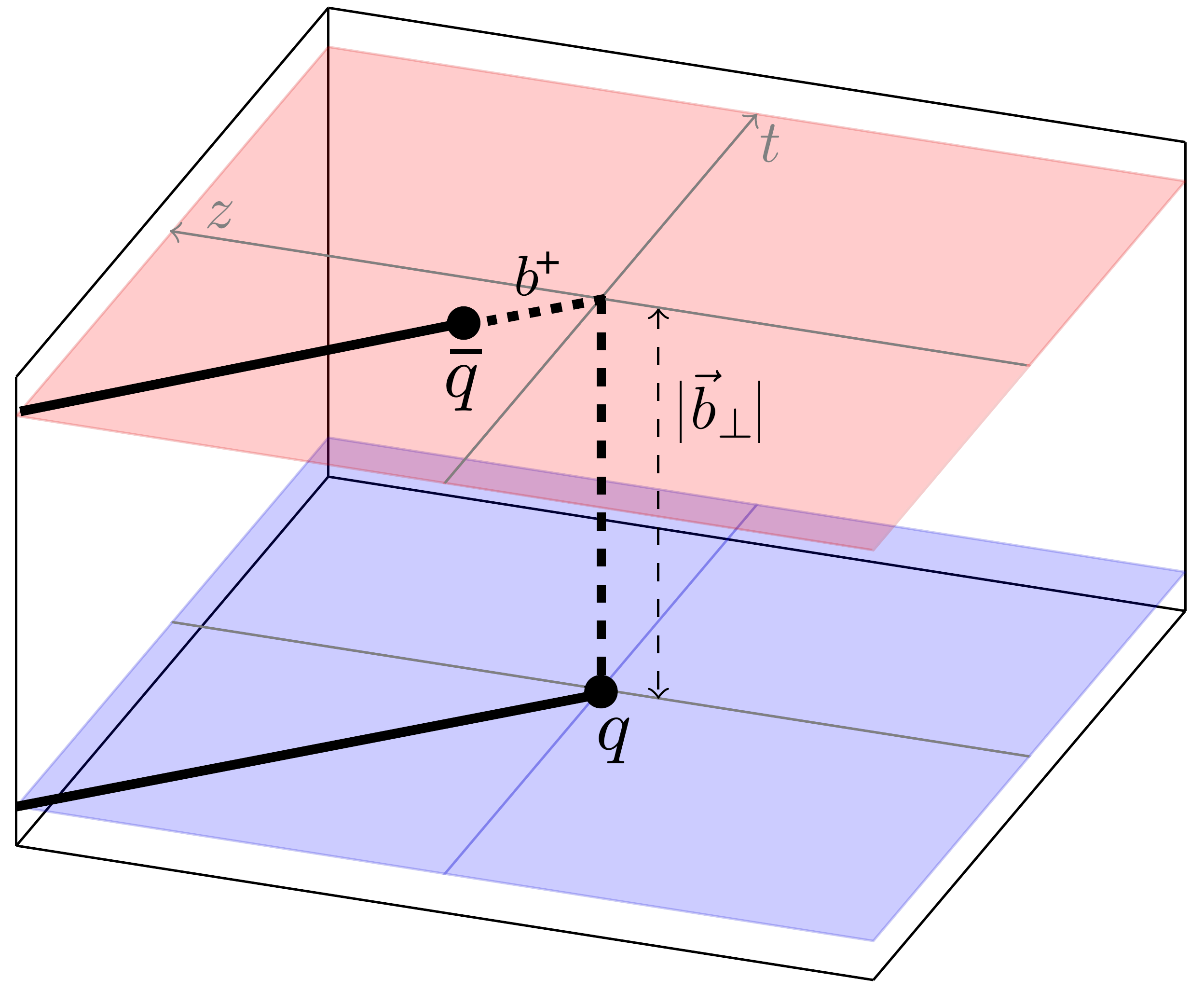}
    \qquad\qquad
	\includegraphics[width=0.35\textwidth]{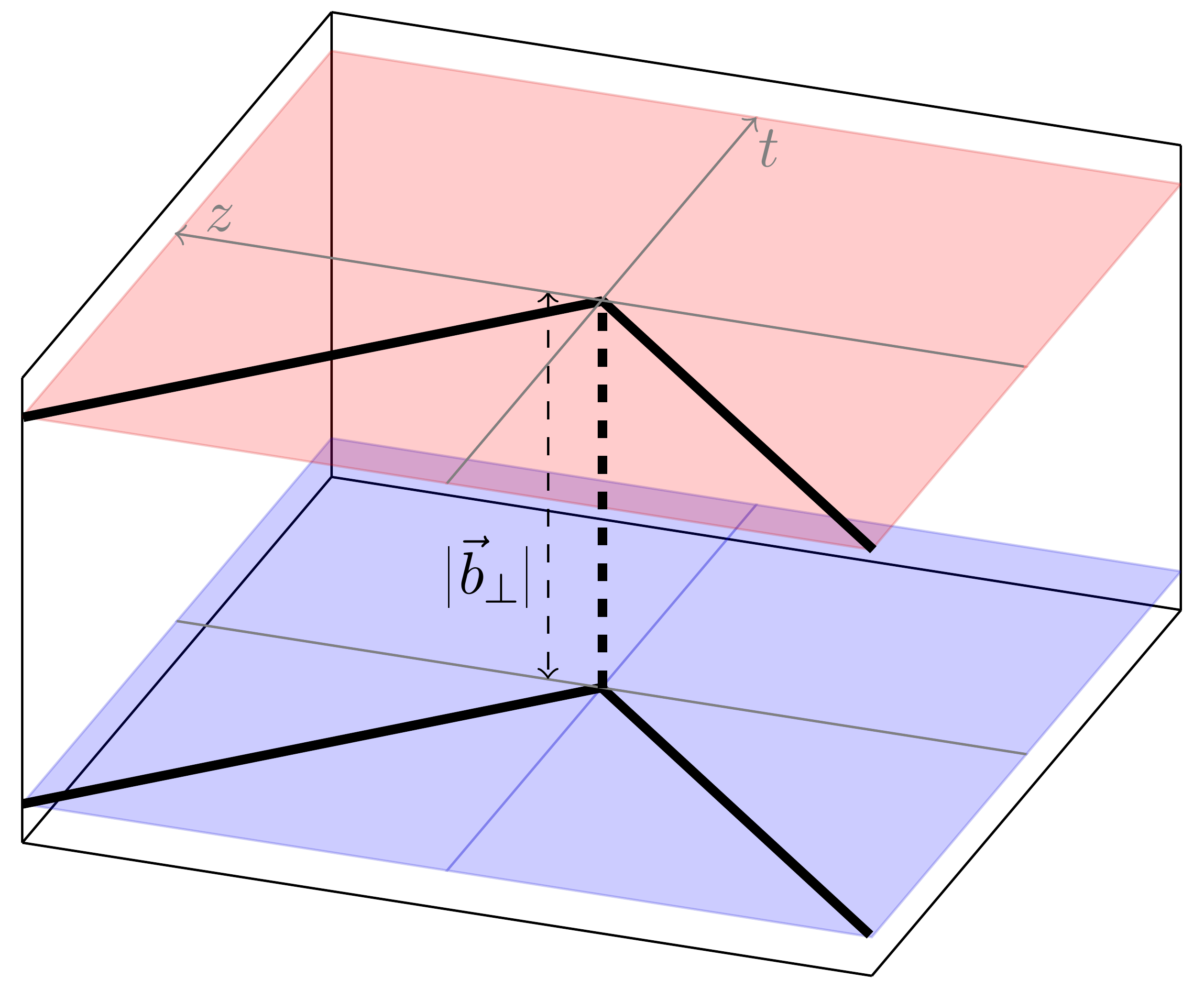}\\[-5pt]
	\raisebox{3.5cm}{ (a) \hspace{7cm} (b) \hspace{7cm}}
	\vspace{-3.5cm}
	\caption{Illustration of the Wilson line structure of (a) the $n$-collinear beam function $B_{q}$
		and (b) the soft function $S^q$, defined in \eqs{beam}{soft}.
		The Wilson lines (solid) extend to infinity in the directions indicated and are joined there by transverse Wilson lines. The $\tau$ dependence that regulates singularities from these Wilson lines is not shown.
		Adapted from \mycite{Li:2016axz}.}
	\label{fig:wilsonlines}
\end{figure*}

\begin{widetext}
In general $B_{q}$ involves a matrix element with an energetic hadron moving along the $n$ direction, and encodes the effect of collinear radiation associated to the hadron $h$ of momentum $P$, and $S^q(b_T,\epsilon,\tau)$ is a bare soft vacuum matrix element which encodes the effect of soft interactions between the incoming partons. Their precise definitions depend on the regulator $\tau$, which we leave implicit for simplicity,
\begin{align} \label{eq:beam}
 B_{q}(x,\bt,\eps,\tau,\zeta) &= \int\frac{\df b^+}{4\pi} e^{-\img \frac12 b^+ (x P^-)}
 \Bigl< h(P) \Bigr|  \Bigl[ \bar q(b^\mu)
 W(b^\mu) \frac{\gamma^-}{2}
 W_{T}\bigl(-\infty\bn;\vec b_T,\vec 0_T\bigr)
 W^\dagger(0)  q(0) \Bigr]_\tau \Bigl| h(P) \Bigr>
,\\  \label{eq:soft}
 S^q(b_T,\eps,\tau) &= \frac{1}{N_c} \bigl< 0 \bigr| {\rm Tr} \bigl[ S^\dagger_n(\bt) S_\bn(\bt)
   S_{T}(-\infty \bn;\vec b_T,\vec 0_T)
 S^\dagger_\bn(\vec 0_T) S_n(\vec 0_T)
 S_{T}^\dagger\bigl(-\infty n;\vec b_T,\vec 0_T\bigr) \bigr]_\tau
 \bigl|0 \bigr>
\,.\end{align}
\end{widetext}
We use the lightcone coordinates $b^\pm = b^0 \mp b^z$, such that $b^\mu = b^+ \bn^\mu/2 + b_T^\mu$.
The Wilson lines appearing in \eqs{beam}{soft} are defined as path-ordered exponentials,
{\allowdisplaybreaks
\begin{align} \label{eq:Wilson_lines}
 W(x^\mu) &= P \exp\biggl[ -\img g \int_{-\infty}^0 \df s\, \bn \cdot \cA(x^\mu + s \bn^\mu) \biggr]
\,,\nn\\
 S_n(x^\mu) &= P \exp\biggl[ -\img g \int_{-\infty}^0 \df s\, n \cdot \cA(x^\mu + s n^\mu) \biggr]
\,,\nn\\
 W_{T}(x^\mu;\vec b_T,\vec 0_T) &
 = P \exp\left[ \img g \int_{\vec 0_T}^{\vec b_T} \df \vec s_T \cdot \vec \cA_T(x^\mu + s_T^\mu) \right]
 \nn\\*&
 = S_{T}(x^\mu;\vec b_T,\vec 0_T)
\,.\end{align}
}%
For clarity, we use a different notation for soft Wilson lines $S$ and collinear Wilson lines $W$.
The Wilson line paths in \eqs{beam}{soft} are illustrated in \fig{wilsonlines}.

The dependence of the TMDPDF on the scales $\mu$ and $\zeta$ is governed by the renormalization group equations
\begin{align} \label{eq:TMD_RGEs}
 \mu \frac{\df}{\df \mu} &f^\TMD_i(x, \bt, \mu, \zeta) = \gamma_\mu^i(\mu,\zeta) f^\TMD_i(x, \bt, \mu, \zeta)
\,,\nn\\
 \zeta \frac{\df}{\df \zeta} &f^\TMD_i(x, \bt, \mu, \zeta) = \frac12 \gamma_\zeta^i(\mu, b_T) f^\TMD_i(x, \bt, \mu, \zeta)
\,,\nn\\
 \mu \frac{\df}{\df\mu} &\gamma_\zeta^i(\mu,b_T) = 2 \zeta \frac{\df}{\df\zeta} \gamma_\mu^i(\mu,\zeta) = - 2\GammaC^i[\as(\mu)]
\,,\end{align}
where the second equation is the Collins-Soper equation \cite{Collins:1981va,Collins:1981uk}.
It can also be written as a convolution in momentum space \cite{Collins:1981va}, where its solution is more complicated \cite{Ebert:2016gcn}.
The combined solution to \eq{TMD_RGEs} yields the evolution in \eq{evolution}, where we have chosen a specific path in the $(\mu,\zeta)$ plane, but one is free to choose any path connecting $(\mu_0,\zeta_0)\to(\mu,\zeta)$ since the last equation in \eq{TMD_RGEs} ensures path independence (see also \mycite{Scimemi:2018xaf}).

The subscripts on the anomalous dimensions $\gamma_\mu^i$ and $\gamma_\zeta^i$ label the scale evolution they govern.
Their all-order forms are given by
\begin{align} \label{eq:anom_dims}
 \gamma_\mu^i(\mu,\zeta) &= \GammaC^i[\as(\mu)] \ln\frac{\mu^2}{\zeta} + \gamma_\mu^i[\as(\mu)]
\,,\\\nn
 \gamma_\zeta^i(\mu,b_T) &= -2 \int_{1/b_T}^\mu \frac{\df\mu'}{\mu'} \GammaC^i[\as(\mu')] + \gamma_\zeta^i[\as(1/b_T)]
\,.\end{align}
They both have a piece governed by the cusp anomalous dimension $\GammaC^i[\as]$, and a noncusp piece $\gamma^i[\as]$.
Both anomalous dimension differ for quarks, $i=q$, and gluons, $i=g$, but are independent of the choice of hadron state and the light quark flavor in the operator in \eq{beam} (for $b$-quarks see \mycite{Pietrulewicz:2017gxc}).
For $\gamma_\zeta^i$ this follows because 
\begin{align}
\gamma_\zeta^i &= 2 \frac{\df\ln f_i^\TMD}{\df\ln \zeta}  = 2 \frac{\df\ln B_i }{\df\ln \zeta} 
= - \frac{\df\ln B_i}{\df (1/\tau)}  =   \frac{\df \ln \Delta_S^i}{\df (1/\tau)} \,,
\end{align} 
and $\Delta_S^i$ does not depend on the hadron state. 

\eq{anom_dims} clearly shows that  if $b_T \sim \LQCD^{-1}$ then $\gamma_\zeta^i(\mu,b_T)$ has an intrinsically nonperturbative component.
Once $\gamma_\zeta^i(\mu,b_T)$ is determined at a scale $\mu_0 \gg \LQCD$,
it can be perturbatively determined at any scale $\mu\gg \LQCD$ via
\begin{align}
 \gamma_\zeta^i(\mu,b_T) = \gamma_\zeta^i(\mu_0,b_T) - 2\int_{\mu_0}^\mu  \frac{\df\mu'}{\mu'} \GammaC^i[\as(\mu')]
\,.\end{align}
The result for $\gamma_\zeta^i(\mu_0,b_T)$ is known to 3-loop order for perturbative $b_T$~\cite{Li:2016axz,Li:2016ctv,Vladimirov:2016dll}.
The focus of this paper is to determine  $\gamma_\zeta^i(\mu_0,b_T)$ nonperturbatively, which can be used for the evolution even when $b_T\sim \LQCD^{-1}$.

\vspace{0.2cm}
\section{TMDPDFs on the Lattice}
\label{sec:quasi_tmd}
\vspace{-0.2cm}

While lattice QCD provides a practical tool for first-principle calculations of nonperturbative quantities, it has long been challenging to compute lightcone correlators on the Euclidean lattice due to their real-time dependence.
LaMET has been proposed to overcome this hurdle by relating the light-cone correlator to an equal-time correlator in a highly boosted hadron state \cite{Ji:2013dva,Ji:2014gla}.
The latter can be calculated on lattice, and can then be matched onto the corresponding lightcone matrix elements through a systematic expansion in the hadron momentum $P^z$,
as proven in \mycites{Ma:2014jla,Ma:2017pxb,Izubuchi:2018srq}.

\begin{figure*}[t]
\centering
\includegraphics[width=0.35\textwidth]{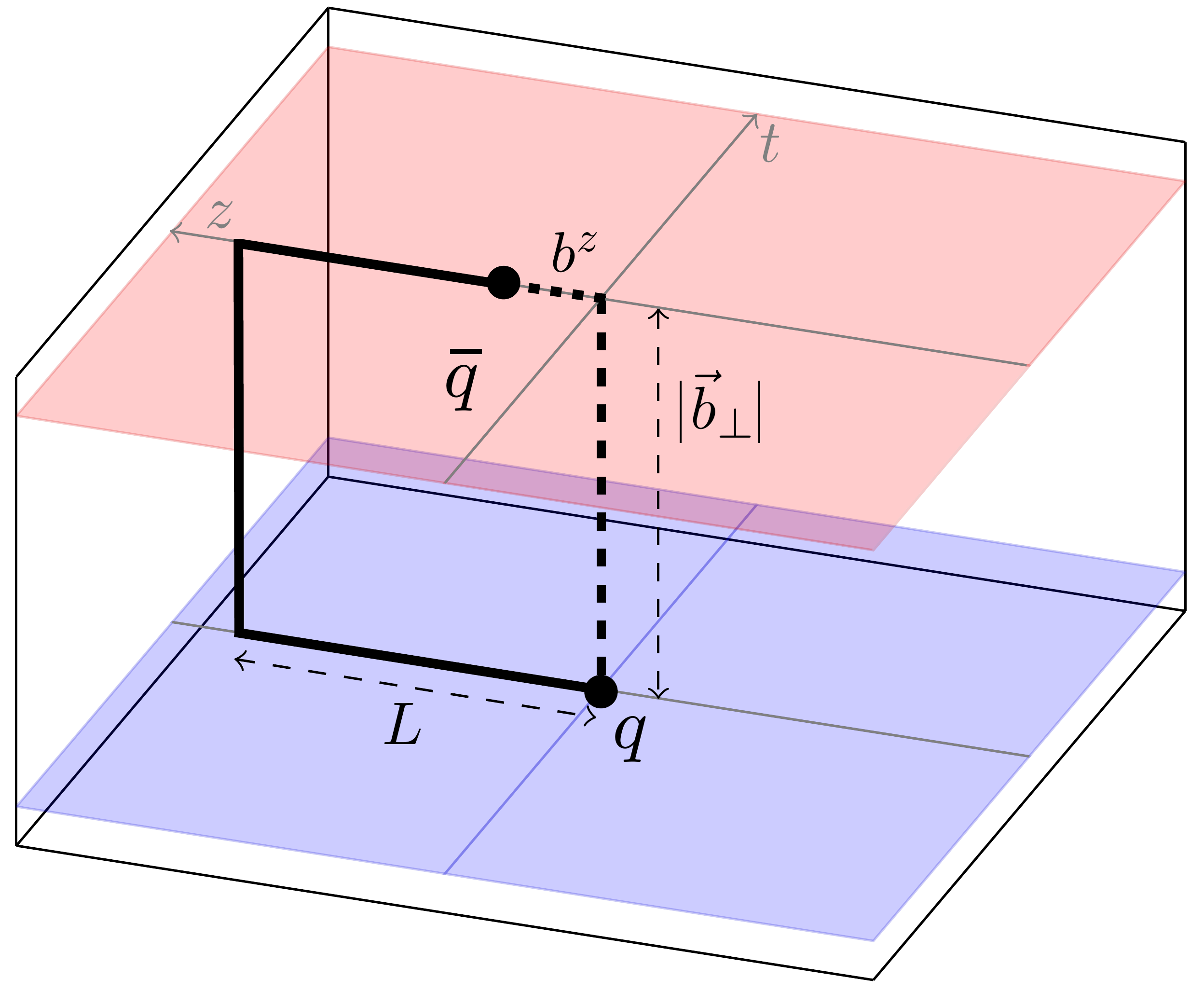} \qquad\qquad
\includegraphics[width=0.3\textwidth]{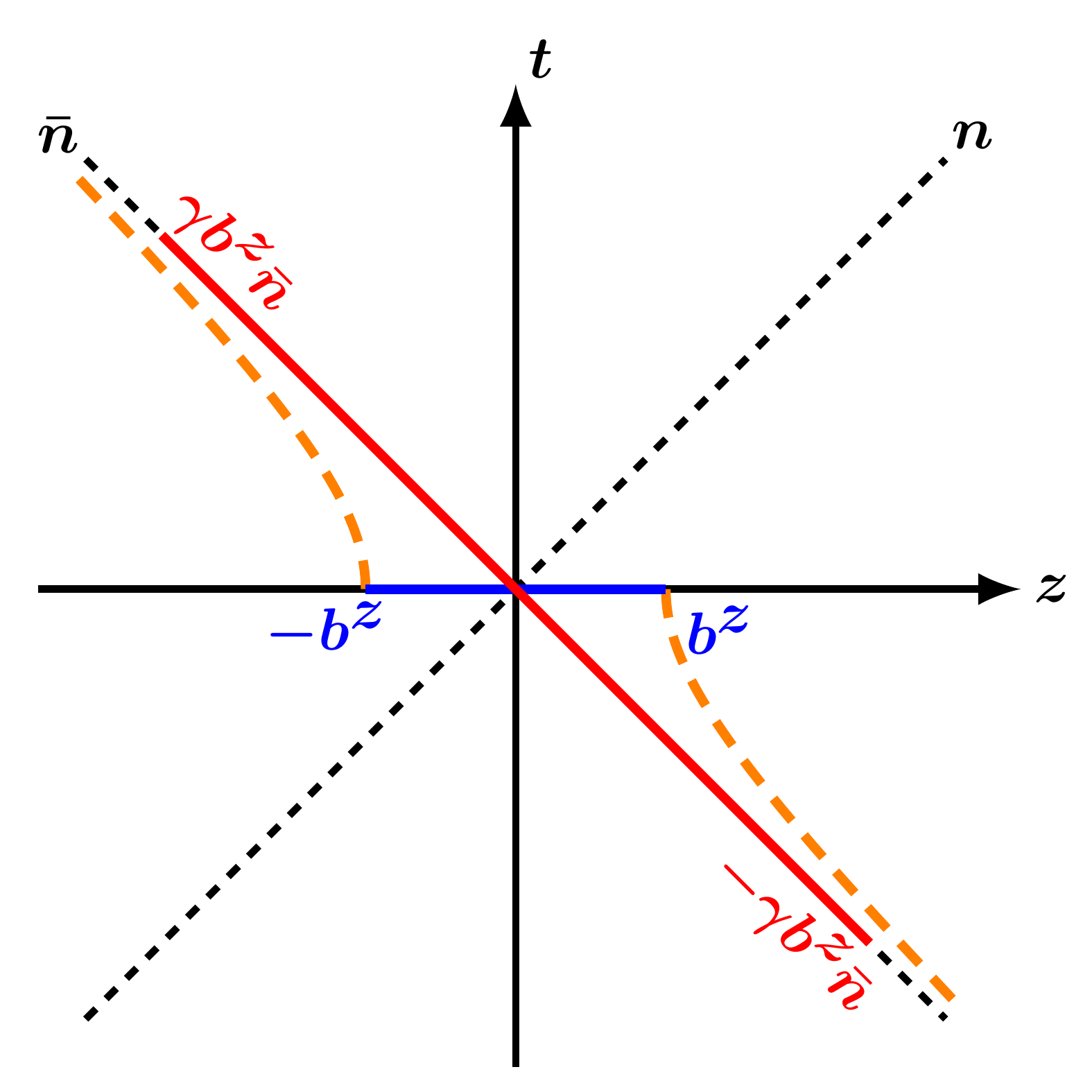}
\\[-5pt]
\raisebox{3.5cm}{ (a) \hspace{7cm} (b) \hspace{7cm}}
\vspace{-3.5cm}
 \caption{(a): Illustration of the Wilson line structure of the quasi beam function $\tilde B_q$ in \eq{qbeam}.
 (b): Behavior of a longitudinal separation $b^z$ (blue solid) under a Lorentz boost along the $z$ direction (orange dotted), and its approximate limit $-\gamma b^z \bn$.}
\label{fig:qbeam}
\end{figure*}

\begin{widetext}
Using the LaMET approach, it has been suggested to calculate TMDPDFs in a similar fashion \cite{Ji:2014hxa,Ji:2018hvs,Ebert:2019okf}. 
An earlier and related approach used in lattice calculations is to exploit Lorentz invariance for the spacelike correlator~\cite{Musch:2010ka,Musch:2011er}.
To begin with, we define a bare \emph{quasi beam function} in position space as
\begin{align} \label{eq:qbeam}
 \tilde B_{q}(b^z, \bt, a,L,P^z)
 =
 \Bigl< h(P) \Big| &\bar q(b^\mu) W_{\hat z}(b^\mu;L-b^z) \frac{\Gamma}{2}
 W_T(L \hat z; \bt, \vec{0}_T) W^\dagger_{\hat z}(0;L) q(0) \Big| h(P) \Bigr>
\,,\end{align}
where $b^\mu = (0, \bt, b^z)$ and $a$ denotes the lattice spacing which acts as an UV regulator, but for simplicity we stick to continuum notation for the fields.
In \eq{qbeam}, one has either $\Gamma=\gamma^0$ or $\Gamma=\gamma^z$, as both can be boosted onto $\gamma^-$.  Just like for the quasi-PDF the choice of $\Gamma=\gamma^0$ might be preferred on the lattice to avoid operator mixing~\cite{Constantinou:2017sej,Green:2017xeu,Chen:2017mie}.
\end{widetext}

Compared to \eq{beam}, in \eq{qbeam} the lightcone Wilson lines are replaced by purely spatial Wilson lines
of length $L$ because of the finite lattice size,
\begin{align} \label{eq:coll_Wilson_L}
 W_{\hat z}(x^\mu;L) &= P \exp\left[ \img g \int_{L}^0 \df s \, \cA^z(x^\mu + s \hat z) \right]
\,.\end{align}
This also regulates the analog of rapidity divergences in $\tilde B_q$,
as has been shown explicitly in \mycites{Ji:2018hvs,Ebert:2019okf}, and thus $L$, in part, takes the role of the rapidity regulator $\tau$ in \eq{beam}.
The inclusion of transverse gauge links ensures gauge invariance \cite{Ji:2002aa,Belitsky:2002sm,Idilbi:2010im,GarciaEchevarria:2011md}.
The resulting Wilson line structure of \eq{qbeam} is illustrated in \fig{qbeam}a. The same correlator in \eq{qbeam} has been used in the lattice calculation of ratios of TMDPDFs with $b^z=0$ in \mycites{Musch:2011er,Engelhardt:2015xja,Yoon:2016dyh,Yoon:2017qzo}.

Under a Lorentz boost along the $z$ direction with velocity $v\,{\to}\,1$ and boost parameter $\gamma=1/\sqrt{1-v^2}$, the spatial separation behaves as $\hat z = (0,0,0,1) \to \gamma (-v,0,0,1) \approx -\gamma \bn$.
This boost behavior is illustrated in \fig{qbeam}b.
It is easy to see that by applying this relation to \eq{qbeam}, one recovers \eq{beam} in the limit $v \to 1$.
This suggests that a matching between \eqs{beam}{qbeam} could be possible, similar to the collinear PDF,
up to possible issues from regulating rapidity divergences, as such regulators necessarily break boost invariance, see \mycite{Ebert:2019okf}.

For the soft matrix element defined in \eq{soft}, a simple quasi construction is not possible,
since the Wilson lines involve both lightcone directions $n$ and $\bn$, which would require opposite boosts to be recovered from spatial Wilson lines along the $\pm \hat z$ directions.
A detailed study of this issue is given in \mycite{Ebert:2019okf}.
Due to this issue, we will simply introduce an intrinsically nonperturbative quantity $\tilde\Delta_S^q(b_T,a,L)$ to describe the missing infrared (IR) physics and dependence on $b_T$.

We define the quasi-TMDPDF in the $\MS$ scheme analogous to \eq{tmdpdf} as
\begin{align} \label{eq:qtmdpdf}
 \tilde f_{q}^\TMD(x, \bt,\mu,P^z) = \!\int\!\frac{\df b^z}{2\pi} e^{\img b^z (x P^z)} \tilde f_{q}^\TMD(b^z, \bt,\mu,P^z)
\,, \end{align}
where the $\MS$ position-space quasi-TMDPDF is
\begin{align} \label{eq:qtmdpdf_bz}
 \tilde f^\TMD_q(b^z, \bt,  \mu, P^z)
 &= \tilde Z'(b^z,\mu,\tilde \mu) \tilde Z_{\rm uv}(b^z,\tilde \mu, a)
\\*& \times
 \tilde B_{q}(b^z, \bt, a, L, P^z) \tilde\Delta_S^q(b_T, a, L)
\nn\,.\end{align}
Here the quasi soft factor $\tilde\Delta_S^q$ also serves as a counterterm to cancel $L/b_T$ divergences in $\tilde B_{q}$.
The $\tilde Z_{\rm uv}$ carries out UV renormalization, which could be done nonperturbatively on the lattice, and $\tilde \mu$ denotes any renormalization scales this introduces. The conversion factor $\tilde Z'$ converts the result into the $\MS$ scheme at the scale $\mu$, which necessarily is perturbative. On the lattice there will be linear power divergences $\sim 1/a$ coming from the spacelike Wilson-line self energies. The quasi soft factor $\tilde\Delta_S^q$ cancels the $L/a$ Wilson line self energy divergences appearing in $\tilde B_{q}$, which leaves only $b^z/a$ divergences to be canceled by $\tilde Z_{\rm uv}$.
One can consider removing these $1/a$ divergences with a counterterm determined from the static quark-antiquark potential as in \mycite{Musch:2010ka}, or with the RI/MOM scheme like in the quasi-PDF~\cite{Constantinou:2017sej,Stewart:2017tvs}, or with the gradient flow method~\cite{Monahan:2016bvm}.

In \eq{qtmdpdf_bz} we suppress the leftover $L$ dependence which vanishes in the physical limit $L\to\infty$. It would be interesting to construct a direct proof of these renormalization properties of the quasi-TMDPDF, along the lines of those for the quasi-PDF in Ref.~\cite{Ji:2017oey,Ishikawa:2017faj}.  Equation~(\ref{eq:qtmdpdf_bz}) has been explicitly verified in perturbation theory at one-loop order, where definitions of $\tilde\Delta_S^q$ that cancel all divergences in $L$ have also been given \cite{Ji:2018hvs,Ebert:2019okf}.

The quasi-TMDPDF defined in \eq{qtmdpdf} is not a boost invariant quantity and thus explicitly depends on the hadron momentum $P^z$, which also plays the role of the variable $\zeta$ in the TMDPDF. We will exploit this $P^z$ dependence for our method to extract $\gamma_\zeta^q$. Importantly, $\tilde\Delta_S^q$ does not depend on $x$ or $P^z$.
It also does not depend on the quark flavor, but differs from $\tilde\Delta_S^g$ for gluons.
Thus, $\tilde\Delta_S^q$ drops out of ratios that are flavor blind, which will be a crucial ingredient to our proposed method, as one does not need to precisely define or calculate $\tilde\Delta_S^q$ on the lattice. Note that we convert $\tilde f_q^\TMD$ to the $\MS$ scheme with $\tilde Z'$ to simplify carrying out the matching onto the TMDPDF, though matching results for other schemes could be considered.

\subsection{Relating quasi-TMDPDF and TMDPDF}

For the collinear PDF, LaMET gives a perturbative matching between the quasi-PDF and PDF, and the same may be true between the quasi beam function and beam function. However, for the full TMDPDF this matching is potentially spoiled by the presence of the soft factors $\Delta_S^q$ and $\tilde\Delta_S^q$ that cannot be related simply through a Lorentz boost. For our purposes, we define $g^S_q(b_T,\mu)$ as the mismatch of the lightlike and quasi soft factors, which we allow to be nonperturbative, and we will exploit the fact that due to its soft origin it is  independent of $x$, $P^z$, and quark flavor. For a flavor nonsinglet channel such as $u{-}d$ the relation between the $\MS$ quasi-TMDPDF and TMDPDF is thus expected to take the form \cite{Ebert:2019okf}
\begin{align} \label{eq:qTMDtoTMD}
 &\tilde f_\ns^\TMD(x, \bt, \mu, P^z)
 = C^\TMD_\ns\bigl(\mu, x P^z\bigr)
  \: g^S_q(b_T,\mu)
\nn\\*&\times
   \exp\biggl[\frac12\gamma_\zeta^q(\mu, b_T) \ln\frac{(2 x P^z)^2}{\zeta}  \biggr]
    f^\TMD_\ns(x, \bt, \mu, \zeta)
\,.\end{align}
We suppress explicit corrections to \eq{qTMDtoTMD} in $b_T/L$, $1/(b_T P^z)$, $1/(P^z L)$ and $\LQCD/P^z$ arising from the finite hadron momentum $P^z$ and finite lattice size. Like for the quasi-PDF~\cite{Chen:2016utp} there are also hadron mass corrections, $M_h/P^z$, which can likely be accounted for exactly.
\eq{qTMDtoTMD} has been explicitly verified at one loop in Ref.~\cite{Ebert:2019okf}.
It involves a perturbative short distance coefficient $C_\ns^\TMD$, which is independent of $b_T$.
It is multiplicative in $x$ space, which is known to hold at least to one-loop order \cite{Ji:2018hvs,Ebert:2019okf},
in contrast to the quasi-PDF, whose matching onto the PDF involves a convolution in $x$~\cite{Xiong:2013bka}.

The exponential in \eq{qTMDtoTMD} contains the nonperturbative Collins-Soper kernel $\gamma_\zeta^q(\mu,b_T)$ that we are after.
It guarantees that $\tilde f_\ns^\TMD$ is independent of $\zeta$ by balancing the Collins-Soper evolution of $f^\TMD_\ns$.
This exponential is $\bt$ dependent and thus can not be included in the short-distance  $C_\ns^\TMD$.
The $\zeta$ dependence in this exponential is balanced by $\tilde\zeta \equiv (2 x P^z)^2$, which is the Collins-Soper scale of \eq{zeta_Collins} away from the lightcone, corresponding with the $z$-momentum of the struck quark. For our use of \eq{qTMDtoTMD} we are considering $\tilde\zeta$ and $\zeta$ to be independent variables. This can be equivalently thought of as having different values of $P^z$ in the quasi-TMDPDF $\tilde f_\ns^\TMD$ and TMDPDF $f_\ns^\TMD$, where the nonperturbative $\gamma_\zeta^q(\mu,b_T)$ is then needed to connect these two different values of $P^z$.  Note that for our purposes any mismatch in the multiplicative constant in $\zeta$ and $\tilde \zeta$, the analogs of $e^{-2y_n}$ in \eq{zeta_Collins}, can be compensated by a change to $g_q^S$, so we can take these $e^{-2y_n}$ constants to be $1$.  

If a definition of $\tilde\Delta_S^q$ can be found which is calculable on the lattice and matches the IR physics of $\Delta_S^q$, then $g^S_q$ would become independent of $b_T$ and calculable in perturbation theory. In this case \eq{qTMDtoTMD} would become a true matching relation between the quasi-TMDPDF and TMDPDF. For our analysis here we will not need to assume   such an operator definition of $\tilde\Delta_S^q$ exists.

Note that our result in \eq{qTMDtoTMD} differs somewhat from the analog in \mycite{Ji:2018hvs}. There the $P^z$ of the quasi-TMDPDF and TMDPDF were taken to be equal, so the exponential term involving $\gamma_\zeta^q$ does not appear. Also they interpreted the $b_T$ dependence of their $g_q^S(b_T,\mu)$ to be short distance without considering its nonperturbative nature for $b_T\sim \LQCD^{-1}$, thus (incorrectly) concluding that \eq{qTMDtoTMD} with a nontrivial $g_q^S(b_T,\mu)$ could still be interpreted as a matching equation.

\subsection{Explicit relation at one loop}

One-loop calculations comparing the quark nonsinglet quasi-TMDPDF and TMDPDF in the $\MS$ scheme have been carried out in \mycites{Ji:2018hvs,Ebert:2019okf} in the limit of large $P^z$ and $L$.
Both employ a spatial soft matrix element $\tilde\Delta_S^q$ obtained by replacing $n\,{\to}\,\hat z$ and $\bn\,{\to}-\!\hat z$ in \eq{soft}.
\mycite{Ebert:2019okf} separately calculates $\tilde B_q$ and $\tilde\Delta_S^q$, allowing one to easily verify the structure of \eq{qTMDtoTMD}.
One obtains
\begin{align} \label{eq:kernels_nlo}
 &C^\TMD_\ns\bigl(\mu, x P^z\bigr)
  = 1 + \frac{\as C_F}{4\pi} C^{(1)}(\mu, x P^z) + \cO(\as^2)
\,,\!\!\\*[3pt] \nn
 &C^{(1)}(\mu, x P^z) = - \ln^2\frac{(2 x P^z)^2}{\mu^2} + 2 \ln\frac{(2 x P^z)^2}{\mu^2}
 - 4 + \frac{\pi^2}{6},
 \\ 
\label{eq:gS}
 &g^S_q(b_T,\mu) = 1 + \frac{\as C_F}{2\pi} \ln\frac{b_T^2 \mu^2}{4 e^{-2\gamma_E}} + \cO(\as^2)
\,.\end{align}
The result of \mycite{Ji:2018hvs} corresponds to combining \eqs{kernels_nlo}{gS} as in \eq{qTMDtoTMD},
and agrees with \mycite{Ebert:2019okf} up the factor $\pi^2/6$ due to a different definition of the $\MS$ scheme.
We and \mycite{Ebert:2019okf} use $\mu^2_{\MS} = 4\pi e^{-\gamma_E} \mu^2_{\rm MS}$, whereas \mycite{Ji:2018hvs} uses $\mu^{2\eps}_{\MS} = (4\pi \mu^2_{\rm MS})^\eps / \Gamma(1-\eps)$.

\section{\texorpdfstring{\boldmath Extracting $\gamma_\zeta$ on Lattice}{Extracting gammazeta on Lattice}}
\label{sec:gamma_zeta}

The relation between the quasi-TMDPDF and TMDPDF in \eq{qTMDtoTMD} allows us to extract $\gamma_\zeta^q(\mu,b_T)$ using lattice QCD.
By considering the ratio of two copies of \eq{qTMDtoTMD} with different hadron momenta $P^z_1 \ne P^z_2$ in the quasi-TMDPDF and the same $\zeta$ in the TMDPDFs, the TMDPDFs and unknown soft contribution $g^S_q$ cancel out, yielding our main result
\begin{widetext}
\begin{align} \label{eq:result}
 \gamma^q_\zeta(\mu, b_T) &
 = \frac{1}{\ln(P^z_1/P^z_2)}
  \ln \frac{C^\TMD_\ns(\mu,x P_2^z)\, \tilde f^\TMD_\ns(x, \bt, \mu, P^z_1)}
           {C^\TMD_\ns(\mu,x P_1^z)\, \tilde f^\TMD_\ns(x, \bt, \mu, P^z_2)}
 \\
 &= \frac{1}{\ln(P^z_1/P^z_2)}
    \ln \frac{C^\TMD_\ns(\mu,x P_2^z)\, \int\! \df b^z\, e^{ib^z xP_1^z} \tilde Z'(b^z,\mu,\tilde \mu) 
    	   \tilde Z_{\rm uv}(b^z,\tilde \mu, a) \tilde B_{\ns}(b^z, \bt, a, L, P_1^z)}
           {C^\TMD_\ns(\mu,x P_1^z)\, \int\! \df b^z\, e^{ib^z xP_2^z} \tilde Z'(b^z,\mu,\tilde \mu) 
           	\tilde Z_{\rm uv}(b^z,\tilde \mu, a) \tilde B_{\ns}(b^z, \bt, a, L, P_2^z)}
\nn\,.\end{align}
\end{widetext}
The first line of \eq{result} employs the quasi-TMDPDFs from \eqs{qtmdpdf}{qtmdpdf_bz}, while in the second line
we have explicitly canceled out the soft factors $\Delta^q_S(b_T,a,L)$ to express the result entirely in terms of quasi beam functions and renormalization factors. In the second line the divergences from $L\to\infty$ cancel in the ratio.

It is important to note that $\gamma_\zeta^q$ is independent of the choice for $x$, $P_1^z$ and $P_2^z$ on the right hand side of \eq{result}, and any residual dependence on these can thus be used to study systematic uncertainties.
Due to the universality of $\gamma_\zeta^q$, \eq{result} can be evaluated with any hadron state (such as a pion).

It is currently not clear if the gluon anomalous dimension $\gamma^g_\zeta(\mu,b_T)$ can be obtained in the same manner. The concern is that in the analog of \eq{qtmdpdf_bz} the gluon could mix with the singlet quark, making the cancellation of soft factors problematic since $\tilde\Delta_S^q \ne \tilde\Delta_S^g$.
Also $\gamma^g_\zeta(\mu,b_T)$ can not be obtained from  $\gamma^q_\zeta(\mu,b_T)$ with Casimir scaling, which is violated at $\cO(\as^4)$ for $\Gamma_{\rm cusp}^i$~\cite{Boels:2017skl,Moch:2017uml,Grozin:2017css} and thus does not hold for $\gamma^i_\zeta(\mu,b_T)$ nonperturbatively.

\subsection{Illustration at one loop}

For illustration, we explicitly show that we recover the correct Collins-Soper kernel at one loop.
This requires the ratio of the NLO coefficient \eq{kernels_nlo},
\begin{align} \label{eq:ratio_C}
&\frac{C^\TMD_\ns(\mu,x P_2^z)}{C^\TMD_\ns(\mu,x P_1^z)}
\\\nn&
 = 1 + \frac{\as C_F}{\pi} \ln\frac{P^z_1}{P^z_2} \biggl( \ln\frac{4 x^2 P^z_1 P^z_2}{\mu^2} - 1 \biggr) + \cO(\as^2)
\,,\end{align}
and likewise the ratio of the perturbative quasi-TMDPDFs, calculated with on-shell quark states,
\begin{align} \label{eq:ratio_b}
&~\frac{\tilde f^\TMD_\ns(x, \bt, \mu, P^z_1)}{\tilde f^\TMD_\ns(x, \bt, \mu, P^z_2)}
\\\nn&
 = 1 + \frac{\as C_F}{\pi} \ln\frac{P^z_1}{P^z_2}
   \biggl( -\ln\frac{x^2 P^z_1 P^z_2 b_T^2}{e^{-2\gamma_E}} + 1 \biggr)
   + \cO(\as^2)
\,,\end{align}
which can be obtained from the results given in \app{tmd_nlo}.
Inserting \eqs{ratio_C}{ratio_b} into \eq{result}, we obtain
\begin{align} \label{eq:result_nlo}
 \gamma^q_\zeta(\mu, b_T) &
 = \frac{1}{\ln\frac{P^z_1}{P^z_2}} \ln\biggl[1\! - \frac{\as C_F}{\pi} \ln\frac{P^z_1}{P^z_2} \ln\frac{b_T^2 \mu^2}{4 e^{-2\gamma_E}}
 +\! \cO(\as^2) \biggr]
\nn\\*&
 = - \frac{\as C_F}{\pi} \ln\frac{b_T^2 \mu^2}{4 e^{-2\gamma_E}} + \cO(\as^2)
\,,\end{align}
which is exactly the one-loop anomalous dimension.

\section{Conclusions and Outlook}
\label{sec:conclusions}

In this paper, we have proposed a novel approach to determine  from lattice QCD the nonperturbative anomalous dimension $\gamma^q_\zeta(\mu,b_T)$ governing the Collins-Soper evolution of quark TMDPDFs, given in \eq{result}.
It involves matrix elements of an equal-time operator with boosted hadron states (referred to as quasi beam functions), UV renormalization, and a perturbative short distance kernel $C_\ns^\TMD$. These are taken in a ratio with different hadron momentum, such that soft contributions cancel out.
The nonperturbative contribution to $\gamma^q_\zeta(\mu,b_T)$ is required in order to evolve TMDPDFs, determined at some initial scales from experiment (or perhaps in the future a separate lattice calculation), to the scales appearing in other phenomenological applications.

So far the coefficient $C_\ns^\TMD$ is known up to one-loop order for matching an $\MS$ quasi-TMDPDF to the $\MS$ TMDPDF.
To make use of this result for lattice calculations using nonperturbative renormalization schemes would require explicit computations of the corresponding scheme conversion factor $\tilde Z'$ in \eq{qtmdpdf_bz}. Related examples of nonperturbative renormalization are \mycites{Hagler:2009mb,Musch:2010ka} for straight-Wilson line quasi-TMDs,  \mycites{Constantinou:2017sej,Stewart:2017tvs} for application of the RI/MOM scheme to the quasi-PDF, and \mycite{Monahan:2016bvm} for the gradient flow method.
First lattice studies of the equal-time TMD quasi beam function in \eq{qbeam} have been carried out in \mycites{Musch:2011er,Engelhardt:2015xja,Yoon:2016dyh,Yoon:2017qzo}. There, ratios of this quasi beam function with $b^z=0$ and the same $P^z$ were considered, in which case renormalization factors cancel.
Our method to determine $\gamma_\zeta^q$ instead requires to Fourier transform the renormalized results from $b^z$ into $x$ space, and consider a ratio with two different $P^z$ values in order to expose the Collins-Soper kernel. 

A nonperturbative result for $\gamma_\zeta^q(\mu,b_T)$ could also have interesting applications for perturbative $b_T^{-1}\sim q_T\gg \LQCD$. Here it is known both from factorization and renormalon analyses~\cite{Becher:2013iya,Scimemi:2016ffw} that the perturbative series for $\gamma_\zeta^q(\mu,b_T)$ has $\sim b_T^2 \Lambda_{\rm QCD}^2$ power corrections, so one could attempt to obtain the coefficient of this $b_T^2$ power correction from the small $b_T$ limit of a nonperturbative lattice QCD result. In practice this may be difficult due to the need for a $1/(b_T P^z)$ expansion in the relation between the quasi-TMDPDF and TMDPDF in \eq{qTMDtoTMD}.

A possible modification to our method is to consider a boosted quark state rather than a hadron state, which could have computational advantages.
(We thank Will Detmold for this suggestion.)
Since one can only simulate off-shell gauge-fixed quark states with Euclidean momentum $p_E^2 \ge (p^z)^2$ in lattice QCD, this would require a reconsideration of various ingredients in the proposal made here. It would be interesting to pursue this in the future.

\begin{acknowledgments}
We thank John Collins, Will Detmold, Markus Diehl, Michael Engelhardt and Phiala Shanahan for helpful discussions or comments.
This work was supported by the U.S.\ Department of Energy,
Office of Nuclear Physics, from DE-SC0011090
and within the framework of the TMD Topical Collaboration.
I.S.\ was also supported in part by the Simons Foundation through
the Investigator Grant 327942.
M.E.\ was also supported in part by the Alexander von Humboldt Foundation
through a Feodor Lynen Research Fellowship.
\end{acknowledgments}

\appendix

\section{(Quasi) TMDPDF at NLO}
\label{app:tmd_nlo}

Here we summarize explicit results for the bare  TMDPDF $f^\TMD_q(x,\bt,\eps,\zeta)$
and quasi-TMDPDF $\tilde f^\TMD_q(x,\bt,\eps,P^z)$ at one loop. These are
obtained by evaluating the operators \eqs{beam}{qbeam} in an external quark state
of lightlike momentum $p^\mu = (xP^z, 0, 0, xP^z)$, and
combining them with the appropriate soft or quasi soft function as in
\eqs{tmdpdf}{qtmdpdf} without performing the UV renormalization.
In both cases, we use pure dimensional regularization to regulate both UV and IR divergences,
and defined the $\MS$ renormalization scale as $\mu^2_{\MS} = 4\pi e^{-\gamma_E} \mu^2_{\rm MS}$.

\begin{widetext}
The result for the physical TMDPDF is given by
\begin{align} \label{eq:tmdpdf_nlo}
 f_q^{\TMD}(x, \bt, \eps, \zeta)
 = \delta(1-x) &+ \frac{\as C_F}{2\pi} \biggl[ - \biggl(\frac{1}{\eps_{\rm IR}} + L_b\biggr) P_{qq}(x) +  (1-x) \biggr]_+^1\Theta(1-x)\Theta(x)
 \\*\nn&
  + \frac{\as C_F}{2\pi} \delta(1-x) \biggl[ \frac{1}{\eps_{\rm UV}^2}
   + \biggl(\frac{1}{\eps_{\rm UV}} + L_b\biggr) \biggl(\frac{3}{2} + \ln\frac{\mu^2}{\zeta} \biggr)
   - \frac{1}{2} L_b^2  + \frac{1}{2} - \frac{\pi^2}{12} \biggr]
  + \cO(\as^2)
\,.\end{align}
This result is obtained in the rapidity regulators of \mycites{Collins:1350496,Becher:2010tm,Becher:2011dz,Chiu:2011qc,Chiu:2012ir,GarciaEchevarria:2011rb,Li:2016axz},
which only differ by the precise definition of $\zeta$.
For a more detailed comparison of results in the literature see \mycite{Ebert:2019okf}.

For the quasi-TMDPDF, we take the physical limits $P^z\gg b_T^{-1},L\to\infty$, and use $\tilde\Delta_S^q$ obtained by replacing $n\,{\to}\,\hat z$ and $\bn\,{\to}-\!\hat z$ in \eq{soft}.
The result has been first calculated in \mycite{Ji:2018hvs} and was confirmed in \mycite{Ebert:2019okf},
\begin{align} \label{eq:qtmdpdf_nlo}
 \tilde f_{q}^{\TMD}(x,\bt,\eps,P^z)
 = \delta(1-x)
   &+ \frac{\as C_F}{2\pi} \biggl[ - \biggl(\frac{1}{\eps_{\rm IR}} + L_b \biggr) P_{qq}(x) +  (1-x) \biggr]_+^1 \Theta(1-x)\Theta(x)
 \\*\nn&
   + \frac{\as C_F}{2\pi} \delta(1-x) \biggl[ \frac{3}{2} \frac{1}{\eps_{\rm UV}}
   - \frac{1}{2} L_{P^z}^2 - L_{P^z} - \frac{3}{2}
   - \frac{1}{2} L_b^2 + \frac{5}{2} L_b + L_b L_{P^z} \biggr]
 + \cO(\as^2)
\,.\end{align}
\end{widetext}
In \eqs{tmdpdf_nlo}{qtmdpdf_nlo}, $P_{qq}(x) = [(1+x^2)/(1-x)]$ is the quark-quark splitting function,
and $[f(x)]_+^1$ denotes the usual plus distribution such that $\int_0^1 \df x \, [f(x)]_+^1 = 0$.
We also defined
\begin{align}
 L_b = \ln\frac{\mu^2 b_T^2}{4 e^{-2\gamma_E}}
\,,\quad
 L_{P^z} = \ln\frac{\mu^2}{(2 x P^z)^2}
\,,\end{align}
and made the origin of $1/\epsilon$ poles as either IR or UV divergence explicit.

The UV renormalization factors at NLO in the $\MS$ scheme are given by
\begin{align}
 Z_{\rm uv}(\mu, \eps, \zeta) &= 1 - \frac{\as C_F}{2\pi} \biggl[ \frac{1}{\eps^2}
   + \frac{1}{\eps} \biggl(\frac{3}{2} + \ln\frac{\mu^2}{\zeta} \biggr)\biggr]
   + \cO(\as^2)
\,,\nn\\*
 \tilde Z_{\rm uv}(\mu, \eps) &= 1 - \frac{\as C_F}{2\pi} \frac{3}{2\eps}
 + \cO(\as^2)
\,.\end{align}
The difference of \eqs{tmdpdf_nlo}{qtmdpdf_nlo} after UV renormalization
yields the one-loop kernel in \eqs{kernels_nlo}{gS}.

\clearpage
\bibliography{literature}

\begin{thebibliography}{76}%
\makeatletter
\providecommand \@ifxundefined [1]{%
 \@ifx{#1\undefined}
}%
\providecommand \@ifnum [1]{%
 \ifnum #1\expandafter \@firstoftwo
 \else \expandafter \@secondoftwo
 \fi
}%
\providecommand \@ifx [1]{%
 \ifx #1\expandafter \@firstoftwo
 \else \expandafter \@secondoftwo
 \fi
}%
\providecommand \natexlab [1]{#1}%
\providecommand \enquote  [1]{``#1''}%
\providecommand \bibnamefont  [1]{#1}%
\providecommand \bibfnamefont [1]{#1}%
\providecommand \citenamefont [1]{#1}%
\providecommand \href@noop [0]{\@secondoftwo}%
\providecommand \href [0]{\begingroup \@sanitize@url \@href}%
\providecommand \@href[1]{\@@startlink{#1}\@@href}%
\providecommand \@@href[1]{\endgroup#1\@@endlink}%
\providecommand \@sanitize@url [0]{\catcode `\\12\catcode `\$12\catcode
  `\&12\catcode `\#12\catcode `\^12\catcode `\_12\catcode `\%12\relax}%
\providecommand \@@startlink[1]{}%
\providecommand \@@endlink[0]{}%
\providecommand \url  [0]{\begingroup\@sanitize@url \@url }%
\providecommand \@url [1]{\endgroup\@href {#1}{\urlprefix }}%
\providecommand \urlprefix  [0]{URL }%
\providecommand \Eprint [0]{\href }%
\providecommand \doibase [0]{http://dx.doi.org/}%
\providecommand \selectlanguage [0]{\@gobble}%
\providecommand \bibinfo  [0]{\@secondoftwo}%
\providecommand \bibfield  [0]{\@secondoftwo}%
\providecommand \translation [1]{[#1]}%
\providecommand \BibitemOpen [0]{}%
\providecommand \bibitemStop [0]{}%
\providecommand \bibitemNoStop [0]{.\EOS\space}%
\providecommand \EOS [0]{\spacefactor3000\relax}%
\providecommand \BibitemShut  [1]{\csname bibitem#1\endcsname}%
\let\auto@bib@innerbib\@empty
\bibitem [{\citenamefont {H{\"a}gler}\ \emph {et~al.}(2009)\citenamefont
  {H{\"a}gler}, \citenamefont {Musch}, \citenamefont {Negele},\ and\
  \citenamefont {Sch{\"a}fer}}]{Hagler:2009mb}%
  \BibitemOpen
  \bibfield  {author} {\bibinfo {author} {\bibfnamefont {P.}~\bibnamefont
  {H{\"a}gler}}, \bibinfo {author} {\bibfnamefont {B.~U.}\ \bibnamefont
  {Musch}}, \bibinfo {author} {\bibfnamefont {J.~W.}\ \bibnamefont {Negele}}, \
  and\ \bibinfo {author} {\bibfnamefont {A.}~\bibnamefont {Sch{\"a}fer}},\
  }\href@noop {} {\bibfield  {journal} {\bibinfo  {journal} {EPL}\ }\textbf
  {\bibinfo {volume} {88}},\ \bibinfo {pages} {61001} (\bibinfo {year}
  {2009})},\ \Eprint {http://arxiv.org/abs/0908.1283} {arXiv:0908.1283
  [hep-lat]} \BibitemShut {NoStop}%
\bibitem [{\citenamefont {Musch}\ \emph {et~al.}(2011)\citenamefont {Musch},
  \citenamefont {H{\"a}gler}, \citenamefont {Negele},\ and\ \citenamefont
  {Sch{\"a}fer}}]{Musch:2010ka}%
  \BibitemOpen
  \bibfield  {author} {\bibinfo {author} {\bibfnamefont {B.~U.}\ \bibnamefont
  {Musch}}, \bibinfo {author} {\bibfnamefont {P.}~\bibnamefont {H{\"a}gler}},
  \bibinfo {author} {\bibfnamefont {J.~W.}\ \bibnamefont {Negele}}, \ and\
  \bibinfo {author} {\bibfnamefont {A.}~\bibnamefont {Sch{\"a}fer}},\
  }\href@noop {} {\bibfield  {journal} {\bibinfo  {journal} {Phys. Rev.}\
  }\textbf {\bibinfo {volume} {D83}},\ \bibinfo {pages} {094507} (\bibinfo
  {year} {2011})},\ \Eprint {http://arxiv.org/abs/1011.1213} {arXiv:1011.1213
  [hep-lat]} \BibitemShut {NoStop}%
\bibitem [{\citenamefont {Musch}\ \emph {et~al.}(2012)\citenamefont {Musch},
  \citenamefont {H{\"a}gler}, \citenamefont {Engelhardt}, \citenamefont
  {Negele},\ and\ \citenamefont {Sch{\"a}fer}}]{Musch:2011er}%
  \BibitemOpen
  \bibfield  {author} {\bibinfo {author} {\bibfnamefont {B.~U.}\ \bibnamefont
  {Musch}}, \bibinfo {author} {\bibfnamefont {P.}~\bibnamefont {H{\"a}gler}},
  \bibinfo {author} {\bibfnamefont {M.}~\bibnamefont {Engelhardt}}, \bibinfo
  {author} {\bibfnamefont {J.~W.}\ \bibnamefont {Negele}}, \ and\ \bibinfo
  {author} {\bibfnamefont {A.}~\bibnamefont {Sch{\"a}fer}},\ }\href@noop {}
  {\bibfield  {journal} {\bibinfo  {journal} {Phys. Rev.}\ }\textbf {\bibinfo
  {volume} {D85}},\ \bibinfo {pages} {094510} (\bibinfo {year} {2012})},\
  \Eprint {http://arxiv.org/abs/1111.4249} {arXiv:1111.4249 [hep-lat]}
  \BibitemShut {NoStop}%
\bibitem [{\citenamefont {Engelhardt}\ \emph {et~al.}(2016)\citenamefont
  {Engelhardt}, \citenamefont {H{\"a}gler}, \citenamefont {Musch},
  \citenamefont {Negele},\ and\ \citenamefont
  {Sch{\"a}fer}}]{Engelhardt:2015xja}%
  \BibitemOpen
  \bibfield  {author} {\bibinfo {author} {\bibfnamefont {M.}~\bibnamefont
  {Engelhardt}}, \bibinfo {author} {\bibfnamefont {P.}~\bibnamefont
  {H{\"a}gler}}, \bibinfo {author} {\bibfnamefont {B.}~\bibnamefont {Musch}},
  \bibinfo {author} {\bibfnamefont {J.}~\bibnamefont {Negele}}, \ and\ \bibinfo
  {author} {\bibfnamefont {A.}~\bibnamefont {Sch{\"a}fer}},\ }\href@noop {}
  {\bibfield  {journal} {\bibinfo  {journal} {Phys. Rev.}\ }\textbf {\bibinfo
  {volume} {D93}},\ \bibinfo {pages} {054501} (\bibinfo {year} {2016})},\
  \Eprint {http://arxiv.org/abs/1506.07826} {arXiv:1506.07826 [hep-lat]}
  \BibitemShut {NoStop}%
\bibitem [{\citenamefont {Yoon}\ \emph {et~al.}(2015)\citenamefont {Yoon},
  \citenamefont {Bhattacharya}, \citenamefont {Engelhardt}, \citenamefont
  {Green}, \citenamefont {Gupta}, \citenamefont {H{\"a}gler}, \citenamefont
  {Musch}, \citenamefont {Negele}, \citenamefont {Pochinsky},\ and\
  \citenamefont {Syritsyn}}]{Yoon:2016dyh}%
  \BibitemOpen
  \bibfield  {author} {\bibinfo {author} {\bibfnamefont {B.}~\bibnamefont
  {Yoon}}, \bibinfo {author} {\bibfnamefont {T.}~\bibnamefont {Bhattacharya}},
  \bibinfo {author} {\bibfnamefont {M.}~\bibnamefont {Engelhardt}}, \bibinfo
  {author} {\bibfnamefont {J.}~\bibnamefont {Green}}, \bibinfo {author}
  {\bibfnamefont {R.}~\bibnamefont {Gupta}}, \bibinfo {author} {\bibfnamefont
  {P.}~\bibnamefont {H{\"a}gler}}, \bibinfo {author} {\bibfnamefont
  {B.}~\bibnamefont {Musch}}, \bibinfo {author} {\bibfnamefont
  {J.}~\bibnamefont {Negele}}, \bibinfo {author} {\bibfnamefont
  {A.}~\bibnamefont {Pochinsky}}, \ and\ \bibinfo {author} {\bibfnamefont
  {S.}~\bibnamefont {Syritsyn}},\ }in\ \href@noop {} {\emph {\bibinfo
  {booktitle} {{Proceedings, 33rd International Symposium on Lattice Field
  Theory (Lattice 2015): Kobe, Japan, July 14-18, 2015}}}},\ \bibinfo
  {organization} {SISSA}\ (\bibinfo  {publisher} {SISSA},\ \bibinfo {year}
  {2015})\ \Eprint {http://arxiv.org/abs/1601.05717} {arXiv:1601.05717
  [hep-lat]} \BibitemShut {NoStop}%
\bibitem [{\citenamefont {Yoon}\ \emph {et~al.}(2017)\citenamefont {Yoon},
  \citenamefont {Engelhardt}, \citenamefont {Gupta}, \citenamefont
  {Bhattacharya}, \citenamefont {Green}, \citenamefont {Musch}, \citenamefont
  {Negele}, \citenamefont {Pochinsky}, \citenamefont {Sch{\"a}fer},\ and\
  \citenamefont {Syritsyn}}]{Yoon:2017qzo}%
  \BibitemOpen
  \bibfield  {author} {\bibinfo {author} {\bibfnamefont {B.}~\bibnamefont
  {Yoon}}, \bibinfo {author} {\bibfnamefont {M.}~\bibnamefont {Engelhardt}},
  \bibinfo {author} {\bibfnamefont {R.}~\bibnamefont {Gupta}}, \bibinfo
  {author} {\bibfnamefont {T.}~\bibnamefont {Bhattacharya}}, \bibinfo {author}
  {\bibfnamefont {J.~R.}\ \bibnamefont {Green}}, \bibinfo {author}
  {\bibfnamefont {B.~U.}\ \bibnamefont {Musch}}, \bibinfo {author}
  {\bibfnamefont {J.~W.}\ \bibnamefont {Negele}}, \bibinfo {author}
  {\bibfnamefont {A.~V.}\ \bibnamefont {Pochinsky}}, \bibinfo {author}
  {\bibfnamefont {A.}~\bibnamefont {Sch{\"a}fer}}, \ and\ \bibinfo {author}
  {\bibfnamefont {S.~N.}\ \bibnamefont {Syritsyn}},\ }\href@noop {} {\bibfield
  {journal} {\bibinfo  {journal} {Phys. Rev.}\ }\textbf {\bibinfo {volume}
  {D96}},\ \bibinfo {pages} {094508} (\bibinfo {year} {2017})},\ \Eprint
  {http://arxiv.org/abs/1706.03406} {arXiv:1706.03406 [hep-lat]} \BibitemShut
  {NoStop}%
\bibitem [{\citenamefont {Ji}(2013)}]{Ji:2013dva}%
  \BibitemOpen
  \bibfield  {author} {\bibinfo {author} {\bibfnamefont {X.}~\bibnamefont
  {Ji}},\ }\href@noop {} {\bibfield  {journal} {\bibinfo  {journal} {Phys. Rev.
  Lett.}\ }\textbf {\bibinfo {volume} {110}},\ \bibinfo {pages} {262002}
  (\bibinfo {year} {2013})},\ \Eprint {http://arxiv.org/abs/1305.1539}
  {arXiv:1305.1539 [hep-ph]} \BibitemShut {NoStop}%
\bibitem [{\citenamefont {Ji}(2014)}]{Ji:2014gla}%
  \BibitemOpen
  \bibfield  {author} {\bibinfo {author} {\bibfnamefont {X.}~\bibnamefont
  {Ji}},\ }\href@noop {} {\bibfield  {journal} {\bibinfo  {journal} {Sci. China
  Phys. Mech. Astron.}\ }\textbf {\bibinfo {volume} {57}},\ \bibinfo {pages}
  {1407} (\bibinfo {year} {2014})},\ \Eprint {http://arxiv.org/abs/1404.6680}
  {arXiv:1404.6680 [hep-ph]} \BibitemShut {NoStop}%
\bibitem [{\citenamefont {Ji}\ \emph {et~al.}(2015)\citenamefont {Ji},
  \citenamefont {Sun}, \citenamefont {Xiong},\ and\ \citenamefont
  {Yuan}}]{Ji:2014hxa}%
  \BibitemOpen
  \bibfield  {author} {\bibinfo {author} {\bibfnamefont {X.}~\bibnamefont
  {Ji}}, \bibinfo {author} {\bibfnamefont {P.}~\bibnamefont {Sun}}, \bibinfo
  {author} {\bibfnamefont {X.}~\bibnamefont {Xiong}}, \ and\ \bibinfo {author}
  {\bibfnamefont {F.}~\bibnamefont {Yuan}},\ }\href@noop {} {\bibfield
  {journal} {\bibinfo  {journal} {Phys. Rev.}\ }\textbf {\bibinfo {volume}
  {D91}},\ \bibinfo {pages} {074009} (\bibinfo {year} {2015})},\ \Eprint
  {http://arxiv.org/abs/1405.7640} {arXiv:1405.7640 [hep-ph]} \BibitemShut
  {NoStop}%
\bibitem [{\citenamefont {Ji}\ \emph {et~al.}(2018{\natexlab{a}})\citenamefont
  {Ji}, \citenamefont {Jin}, \citenamefont {Yuan}, \citenamefont {Zhang},\ and\
  \citenamefont {Zhao}}]{Ji:2018hvs}%
  \BibitemOpen
  \bibfield  {author} {\bibinfo {author} {\bibfnamefont {X.}~\bibnamefont
  {Ji}}, \bibinfo {author} {\bibfnamefont {L.-C.}\ \bibnamefont {Jin}},
  \bibinfo {author} {\bibfnamefont {F.}~\bibnamefont {Yuan}}, \bibinfo {author}
  {\bibfnamefont {J.-H.}\ \bibnamefont {Zhang}}, \ and\ \bibinfo {author}
  {\bibfnamefont {Y.}~\bibnamefont {Zhao}},\ }\href@noop {} {\  (\bibinfo
  {year} {2018}{\natexlab{a}})},\ \Eprint {http://arxiv.org/abs/1801.05930}
  {arXiv:1801.05930 [hep-ph]} \BibitemShut {NoStop}%
\bibitem [{\citenamefont {Ebert}\ \emph {et~al.}(2019)\citenamefont {Ebert},
  \citenamefont {Stewart},\ and\ \citenamefont {Zhao}}]{Ebert:2019okf}%
  \BibitemOpen
  \bibfield  {author} {\bibinfo {author} {\bibfnamefont {M.~A.}\ \bibnamefont
  {Ebert}}, \bibinfo {author} {\bibfnamefont {I.~W.}\ \bibnamefont {Stewart}},
  \ and\ \bibinfo {author} {\bibfnamefont {Y.}~\bibnamefont {Zhao}},\
  }\href@noop {} {\  (\bibinfo {year} {2019})},\ \Eprint
  {http://arxiv.org/abs/1901.03685} {arXiv:1901.03685 [hep-ph]} \BibitemShut
  {NoStop}%
\bibitem [{\citenamefont {Collins}\ and\ \citenamefont
  {Soper}(1981)}]{Collins:1981uk}%
  \BibitemOpen
  \bibfield  {author} {\bibinfo {author} {\bibfnamefont {J.~C.}\ \bibnamefont
  {Collins}}\ and\ \bibinfo {author} {\bibfnamefont {D.~E.}\ \bibnamefont
  {Soper}},\ }\href@noop {} {\bibfield  {journal} {\bibinfo  {journal} {Nucl.
  Phys.}\ }\textbf {\bibinfo {volume} {B193}},\ \bibinfo {pages} {381}
  (\bibinfo {year} {1981})},\ \bibinfo {note} {[Erratum: Nucl.
  Phys.B213,545(1983)]}\BibitemShut {NoStop}%
\bibitem [{\citenamefont {Collins}\ and\ \citenamefont
  {Soper}(1982)}]{Collins:1981va}%
  \BibitemOpen
  \bibfield  {author} {\bibinfo {author} {\bibfnamefont {J.~C.}\ \bibnamefont
  {Collins}}\ and\ \bibinfo {author} {\bibfnamefont {D.~E.}\ \bibnamefont
  {Soper}},\ }\href@noop {} {\bibfield  {journal} {\bibinfo  {journal} {Nucl.
  Phys.}\ }\textbf {\bibinfo {volume} {B197}},\ \bibinfo {pages} {446}
  (\bibinfo {year} {1982})}\BibitemShut {NoStop}%
\bibitem [{\citenamefont {Collins}\ \emph
  {et~al.}(1985{\natexlab{a}})\citenamefont {Collins}, \citenamefont {Soper},\
  and\ \citenamefont {Sterman}}]{Collins:1984kg}%
  \BibitemOpen
  \bibfield  {author} {\bibinfo {author} {\bibfnamefont {J.~C.}\ \bibnamefont
  {Collins}}, \bibinfo {author} {\bibfnamefont {D.~E.}\ \bibnamefont {Soper}},
  \ and\ \bibinfo {author} {\bibfnamefont {G.~F.}\ \bibnamefont {Sterman}},\
  }\href@noop {} {\bibfield  {journal} {\bibinfo  {journal} {Nucl. Phys.}\
  }\textbf {\bibinfo {volume} {B250}},\ \bibinfo {pages} {199} (\bibinfo {year}
  {1985}{\natexlab{a}})}\BibitemShut {NoStop}%
\bibitem [{\citenamefont {Collins}(2011)}]{Collins:1350496}%
  \BibitemOpen
  \bibfield  {author} {\bibinfo {author} {\bibfnamefont {J.}~\bibnamefont
  {Collins}},\ }\href@noop {} {\emph {\bibinfo {title} {{Foundations of
  perturbative QCD}}}},\ Cambridge monographs on particle physics, nuclear
  physics, and cosmology\ (\bibinfo  {publisher} {Cambridge Univ. Press},\
  \bibinfo {address} {New York, NY},\ \bibinfo {year} {2011})\BibitemShut
  {NoStop}%
\bibitem [{\citenamefont {Bacchetta}\ \emph {et~al.}(2017)\citenamefont
  {Bacchetta}, \citenamefont {Delcarro}, \citenamefont {Pisano}, \citenamefont
  {Radici},\ and\ \citenamefont {Signori}}]{Bacchetta:2017gcc}%
  \BibitemOpen
  \bibfield  {author} {\bibinfo {author} {\bibfnamefont {A.}~\bibnamefont
  {Bacchetta}}, \bibinfo {author} {\bibfnamefont {F.}~\bibnamefont {Delcarro}},
  \bibinfo {author} {\bibfnamefont {C.}~\bibnamefont {Pisano}}, \bibinfo
  {author} {\bibfnamefont {M.}~\bibnamefont {Radici}}, \ and\ \bibinfo {author}
  {\bibfnamefont {A.}~\bibnamefont {Signori}},\ }\href@noop {} {\bibfield
  {journal} {\bibinfo  {journal} {JHEP}\ }\textbf {\bibinfo {volume} {06}},\
  \bibinfo {pages} {081} (\bibinfo {year} {2017})},\ \Eprint
  {http://arxiv.org/abs/1703.10157} {arXiv:1703.10157 [hep-ph]} \BibitemShut
  {NoStop}%
\bibitem [{\citenamefont {Scimemi}\ and\ \citenamefont
  {Vladimirov}(2018{\natexlab{a}})}]{Scimemi:2017etj}%
  \BibitemOpen
  \bibfield  {author} {\bibinfo {author} {\bibfnamefont {I.}~\bibnamefont
  {Scimemi}}\ and\ \bibinfo {author} {\bibfnamefont {A.}~\bibnamefont
  {Vladimirov}},\ }\href@noop {} {\bibfield  {journal} {\bibinfo  {journal}
  {Eur. Phys. J.}\ }\textbf {\bibinfo {volume} {C78}},\ \bibinfo {pages} {89}
  (\bibinfo {year} {2018}{\natexlab{a}})},\ \Eprint
  {http://arxiv.org/abs/1706.01473} {arXiv:1706.01473 [hep-ph]} \BibitemShut
  {NoStop}%
\bibitem [{\citenamefont {Collins}\ \emph
  {et~al.}(1985{\natexlab{b}})\citenamefont {Collins}, \citenamefont {Soper},\
  and\ \citenamefont {Sterman}}]{Collins:1985ue}%
  \BibitemOpen
  \bibfield  {author} {\bibinfo {author} {\bibfnamefont {J.~C.}\ \bibnamefont
  {Collins}}, \bibinfo {author} {\bibfnamefont {D.~E.}\ \bibnamefont {Soper}},
  \ and\ \bibinfo {author} {\bibfnamefont {G.~F.}\ \bibnamefont {Sterman}},\
  }\href@noop {} {\bibfield  {journal} {\bibinfo  {journal} {Nucl. Phys.}\
  }\textbf {\bibinfo {volume} {B261}},\ \bibinfo {pages} {104} (\bibinfo {year}
  {1985}{\natexlab{b}})}\BibitemShut {NoStop}%
\bibitem [{\citenamefont {Collins}\ \emph {et~al.}(1988)\citenamefont
  {Collins}, \citenamefont {Soper},\ and\ \citenamefont
  {Sterman}}]{Collins:1988ig}%
  \BibitemOpen
  \bibfield  {author} {\bibinfo {author} {\bibfnamefont {J.~C.}\ \bibnamefont
  {Collins}}, \bibinfo {author} {\bibfnamefont {D.~E.}\ \bibnamefont {Soper}},
  \ and\ \bibinfo {author} {\bibfnamefont {G.~F.}\ \bibnamefont {Sterman}},\
  }\href@noop {} {\bibfield  {journal} {\bibinfo  {journal} {Nucl. Phys.}\
  }\textbf {\bibinfo {volume} {B308}},\ \bibinfo {pages} {833} (\bibinfo {year}
  {1988})}\BibitemShut {NoStop}%
\bibitem [{\citenamefont {Collins}\ \emph {et~al.}(1989)\citenamefont
  {Collins}, \citenamefont {Soper},\ and\ \citenamefont
  {Sterman}}]{Collins:1989gx}%
  \BibitemOpen
  \bibfield  {author} {\bibinfo {author} {\bibfnamefont {J.~C.}\ \bibnamefont
  {Collins}}, \bibinfo {author} {\bibfnamefont {D.~E.}\ \bibnamefont {Soper}},
  \ and\ \bibinfo {author} {\bibfnamefont {G.~F.}\ \bibnamefont {Sterman}},\
  }\href@noop {} {\bibfield  {journal} {\bibinfo  {journal} {Adv. Ser. Direct.
  High Energy Phys.}\ }\textbf {\bibinfo {volume} {5}},\ \bibinfo {pages} {1}
  (\bibinfo {year} {1989})},\ \Eprint {http://arxiv.org/abs/hep-ph/0409313}
  {arXiv:hep-ph/0409313 [hep-ph]} \BibitemShut {NoStop}%
\bibitem [{\citenamefont {Diehl}\ \emph {et~al.}(2016)\citenamefont {Diehl},
  \citenamefont {Gaunt}, \citenamefont {Ostermeier}, \citenamefont
  {Pl{\"o}{\ss}l},\ and\ \citenamefont {Sch{\"a}fer}}]{Diehl:2015bca}%
  \BibitemOpen
  \bibfield  {author} {\bibinfo {author} {\bibfnamefont {M.}~\bibnamefont
  {Diehl}}, \bibinfo {author} {\bibfnamefont {J.~R.}\ \bibnamefont {Gaunt}},
  \bibinfo {author} {\bibfnamefont {D.}~\bibnamefont {Ostermeier}}, \bibinfo
  {author} {\bibfnamefont {P.}~\bibnamefont {Pl{\"o}{\ss}l}}, \ and\ \bibinfo
  {author} {\bibfnamefont {A.}~\bibnamefont {Sch{\"a}fer}},\ }\href@noop {}
  {\bibfield  {journal} {\bibinfo  {journal} {JHEP}\ }\textbf {\bibinfo
  {volume} {01}},\ \bibinfo {pages} {076} (\bibinfo {year} {2016})},\ \Eprint
  {http://arxiv.org/abs/1510.08696} {arXiv:1510.08696 [hep-ph]} \BibitemShut
  {NoStop}%
\bibitem [{\citenamefont {Bauer}\ \emph {et~al.}(2000)\citenamefont {Bauer},
  \citenamefont {Fleming},\ and\ \citenamefont {Luke}}]{Bauer:2000ew}%
  \BibitemOpen
  \bibfield  {author} {\bibinfo {author} {\bibfnamefont {C.~W.}\ \bibnamefont
  {Bauer}}, \bibinfo {author} {\bibfnamefont {S.}~\bibnamefont {Fleming}}, \
  and\ \bibinfo {author} {\bibfnamefont {M.~E.}\ \bibnamefont {Luke}},\
  }\href@noop {} {\bibfield  {journal} {\bibinfo  {journal} {Phys. Rev.}\
  }\textbf {\bibinfo {volume} {D63}},\ \bibinfo {pages} {014006} (\bibinfo
  {year} {2000})},\ \Eprint {http://arxiv.org/abs/hep-ph/0005275}
  {arXiv:hep-ph/0005275 [hep-ph]} \BibitemShut {NoStop}%
\bibitem [{\citenamefont {Bauer}\ \emph {et~al.}(2001)\citenamefont {Bauer},
  \citenamefont {Fleming}, \citenamefont {Pirjol},\ and\ \citenamefont
  {Stewart}}]{Bauer:2000yr}%
  \BibitemOpen
  \bibfield  {author} {\bibinfo {author} {\bibfnamefont {C.~W.}\ \bibnamefont
  {Bauer}}, \bibinfo {author} {\bibfnamefont {S.}~\bibnamefont {Fleming}},
  \bibinfo {author} {\bibfnamefont {D.}~\bibnamefont {Pirjol}}, \ and\ \bibinfo
  {author} {\bibfnamefont {I.~W.}\ \bibnamefont {Stewart}},\ }\href@noop {}
  {\bibfield  {journal} {\bibinfo  {journal} {Phys. Rev.}\ }\textbf {\bibinfo
  {volume} {D63}},\ \bibinfo {pages} {114020} (\bibinfo {year} {2001})},\
  \Eprint {http://arxiv.org/abs/hep-ph/0011336} {arXiv:hep-ph/0011336 [hep-ph]}
  \BibitemShut {NoStop}%
\bibitem [{\citenamefont {Bauer}\ and\ \citenamefont
  {Stewart}(2001)}]{Bauer:2001ct}%
  \BibitemOpen
  \bibfield  {author} {\bibinfo {author} {\bibfnamefont {C.~W.}\ \bibnamefont
  {Bauer}}\ and\ \bibinfo {author} {\bibfnamefont {I.~W.}\ \bibnamefont
  {Stewart}},\ }\href@noop {} {\bibfield  {journal} {\bibinfo  {journal} {Phys.
  Lett.}\ }\textbf {\bibinfo {volume} {B516}},\ \bibinfo {pages} {134}
  (\bibinfo {year} {2001})},\ \Eprint {http://arxiv.org/abs/hep-ph/0107001}
  {arXiv:hep-ph/0107001 [hep-ph]} \BibitemShut {NoStop}%
\bibitem [{\citenamefont {Bauer}\ \emph {et~al.}(2002)\citenamefont {Bauer},
  \citenamefont {Pirjol},\ and\ \citenamefont {Stewart}}]{Bauer:2001yt}%
  \BibitemOpen
  \bibfield  {author} {\bibinfo {author} {\bibfnamefont {C.~W.}\ \bibnamefont
  {Bauer}}, \bibinfo {author} {\bibfnamefont {D.}~\bibnamefont {Pirjol}}, \
  and\ \bibinfo {author} {\bibfnamefont {I.~W.}\ \bibnamefont {Stewart}},\
  }\href@noop {} {\bibfield  {journal} {\bibinfo  {journal} {Phys. Rev.}\
  }\textbf {\bibinfo {volume} {D65}},\ \bibinfo {pages} {054022} (\bibinfo
  {year} {2002})},\ \Eprint {http://arxiv.org/abs/hep-ph/0109045}
  {arXiv:hep-ph/0109045 [hep-ph]} \BibitemShut {NoStop}%
\bibitem [{\citenamefont {Becher}\ and\ \citenamefont
  {Neubert}(2011)}]{Becher:2010tm}%
  \BibitemOpen
  \bibfield  {author} {\bibinfo {author} {\bibfnamefont {T.}~\bibnamefont
  {Becher}}\ and\ \bibinfo {author} {\bibfnamefont {M.}~\bibnamefont
  {Neubert}},\ }\href@noop {} {\bibfield  {journal} {\bibinfo  {journal} {Eur.
  Phys. J.}\ }\textbf {\bibinfo {volume} {C71}},\ \bibinfo {pages} {1665}
  (\bibinfo {year} {2011})},\ \Eprint {http://arxiv.org/abs/1007.4005}
  {arXiv:1007.4005 [hep-ph]} \BibitemShut {NoStop}%
\bibitem [{\citenamefont {Becher}\ \emph {et~al.}(2012)\citenamefont {Becher},
  \citenamefont {Neubert},\ and\ \citenamefont {Wilhelm}}]{Becher:2011xn}%
  \BibitemOpen
  \bibfield  {author} {\bibinfo {author} {\bibfnamefont {T.}~\bibnamefont
  {Becher}}, \bibinfo {author} {\bibfnamefont {M.}~\bibnamefont {Neubert}}, \
  and\ \bibinfo {author} {\bibfnamefont {D.}~\bibnamefont {Wilhelm}},\
  }\href@noop {} {\bibfield  {journal} {\bibinfo  {journal} {JHEP}\ }\textbf
  {\bibinfo {volume} {02}},\ \bibinfo {pages} {124} (\bibinfo {year} {2012})},\
  \Eprint {http://arxiv.org/abs/1109.6027} {arXiv:1109.6027 [hep-ph]}
  \BibitemShut {NoStop}%
\bibitem [{\citenamefont {Becher}\ \emph {et~al.}(2013)\citenamefont {Becher},
  \citenamefont {Neubert},\ and\ \citenamefont {Wilhelm}}]{Becher:2012yn}%
  \BibitemOpen
  \bibfield  {author} {\bibinfo {author} {\bibfnamefont {T.}~\bibnamefont
  {Becher}}, \bibinfo {author} {\bibfnamefont {M.}~\bibnamefont {Neubert}}, \
  and\ \bibinfo {author} {\bibfnamefont {D.}~\bibnamefont {Wilhelm}},\
  }\href@noop {} {\bibfield  {journal} {\bibinfo  {journal} {JHEP}\ }\textbf
  {\bibinfo {volume} {05}},\ \bibinfo {pages} {110} (\bibinfo {year} {2013})},\
  \Eprint {http://arxiv.org/abs/1212.2621} {arXiv:1212.2621 [hep-ph]}
  \BibitemShut {NoStop}%
\bibitem [{\citenamefont {Echevarria}\ \emph {et~al.}(2012)\citenamefont
  {Echevarria}, \citenamefont {Idilbi},\ and\ \citenamefont
  {Scimemi}}]{GarciaEchevarria:2011rb}%
  \BibitemOpen
  \bibfield  {author} {\bibinfo {author} {\bibfnamefont {M.~G.}\ \bibnamefont
  {Echevarria}}, \bibinfo {author} {\bibfnamefont {A.}~\bibnamefont {Idilbi}},
  \ and\ \bibinfo {author} {\bibfnamefont {I.}~\bibnamefont {Scimemi}},\
  }\href@noop {} {\bibfield  {journal} {\bibinfo  {journal} {JHEP}\ }\textbf
  {\bibinfo {volume} {07}},\ \bibinfo {pages} {002} (\bibinfo {year} {2012})},\
  \Eprint {http://arxiv.org/abs/1111.4996} {arXiv:1111.4996 [hep-ph]}
  \BibitemShut {NoStop}%
\bibitem [{\citenamefont {Echevarr{\'\i}a}\ \emph {et~al.}(2013)\citenamefont
  {Echevarr{\'\i}a}, \citenamefont {Idilbi},\ and\ \citenamefont
  {Scimemi}}]{Echevarria:2012js}%
  \BibitemOpen
  \bibfield  {author} {\bibinfo {author} {\bibfnamefont {M.~G.}\ \bibnamefont
  {Echevarr{\'\i}a}}, \bibinfo {author} {\bibfnamefont {A.}~\bibnamefont
  {Idilbi}}, \ and\ \bibinfo {author} {\bibfnamefont {I.}~\bibnamefont
  {Scimemi}},\ }\href@noop {} {\bibfield  {journal} {\bibinfo  {journal} {Phys.
  Lett.}\ }\textbf {\bibinfo {volume} {B726}},\ \bibinfo {pages} {795}
  (\bibinfo {year} {2013})},\ \Eprint {http://arxiv.org/abs/1211.1947}
  {arXiv:1211.1947 [hep-ph]} \BibitemShut {NoStop}%
\bibitem [{\citenamefont {Echevarria}\ \emph {et~al.}(2014)\citenamefont
  {Echevarria}, \citenamefont {Idilbi},\ and\ \citenamefont
  {Scimemi}}]{Echevarria:2014rua}%
  \BibitemOpen
  \bibfield  {author} {\bibinfo {author} {\bibfnamefont {M.~G.}\ \bibnamefont
  {Echevarria}}, \bibinfo {author} {\bibfnamefont {A.}~\bibnamefont {Idilbi}},
  \ and\ \bibinfo {author} {\bibfnamefont {I.}~\bibnamefont {Scimemi}},\
  }\href@noop {} {\bibfield  {journal} {\bibinfo  {journal} {Phys. Rev.}\
  }\textbf {\bibinfo {volume} {D90}},\ \bibinfo {pages} {014003} (\bibinfo
  {year} {2014})},\ \Eprint {http://arxiv.org/abs/1402.0869} {arXiv:1402.0869
  [hep-ph]} \BibitemShut {NoStop}%
\bibitem [{\citenamefont {Chiu}\ \emph
  {et~al.}(2012{\natexlab{a}})\citenamefont {Chiu}, \citenamefont {Jain},
  \citenamefont {Neill},\ and\ \citenamefont {Rothstein}}]{Chiu:2012ir}%
  \BibitemOpen
  \bibfield  {author} {\bibinfo {author} {\bibfnamefont {J.-Y.}\ \bibnamefont
  {Chiu}}, \bibinfo {author} {\bibfnamefont {A.}~\bibnamefont {Jain}}, \bibinfo
  {author} {\bibfnamefont {D.}~\bibnamefont {Neill}}, \ and\ \bibinfo {author}
  {\bibfnamefont {I.~Z.}\ \bibnamefont {Rothstein}},\ }\href@noop {} {\bibfield
   {journal} {\bibinfo  {journal} {JHEP}\ }\textbf {\bibinfo {volume} {05}},\
  \bibinfo {pages} {084} (\bibinfo {year} {2012}{\natexlab{a}})},\ \Eprint
  {http://arxiv.org/abs/1202.0814} {arXiv:1202.0814 [hep-ph]} \BibitemShut
  {NoStop}%
\bibitem [{\citenamefont {Collins}\ and\ \citenamefont
  {Rogers}(2017)}]{Collins:2017oxh}%
  \BibitemOpen
  \bibfield  {author} {\bibinfo {author} {\bibfnamefont {J.}~\bibnamefont
  {Collins}}\ and\ \bibinfo {author} {\bibfnamefont {T.~C.}\ \bibnamefont
  {Rogers}},\ }\href@noop {} {\bibfield  {journal} {\bibinfo  {journal} {Phys.
  Rev.}\ }\textbf {\bibinfo {volume} {D96}},\ \bibinfo {pages} {054011}
  (\bibinfo {year} {2017})},\ \Eprint {http://arxiv.org/abs/1705.07167}
  {arXiv:1705.07167 [hep-ph]} \BibitemShut {NoStop}%
\bibitem [{\citenamefont {Catani}\ \emph {et~al.}(2001)\citenamefont {Catani},
  \citenamefont {de~Florian},\ and\ \citenamefont {Grazzini}}]{Catani:2000vq}%
  \BibitemOpen
  \bibfield  {author} {\bibinfo {author} {\bibfnamefont {S.}~\bibnamefont
  {Catani}}, \bibinfo {author} {\bibfnamefont {D.}~\bibnamefont {de~Florian}},
  \ and\ \bibinfo {author} {\bibfnamefont {M.}~\bibnamefont {Grazzini}},\
  }\href@noop {} {\bibfield  {journal} {\bibinfo  {journal} {Nucl. Phys.}\
  }\textbf {\bibinfo {volume} {B596}},\ \bibinfo {pages} {299} (\bibinfo {year}
  {2001})},\ \Eprint {http://arxiv.org/abs/hep-ph/0008184}
  {arXiv:hep-ph/0008184 [hep-ph]} \BibitemShut {NoStop}%
\bibitem [{\citenamefont {Becher}\ and\ \citenamefont
  {Bell}(2012)}]{Becher:2011dz}%
  \BibitemOpen
  \bibfield  {author} {\bibinfo {author} {\bibfnamefont {T.}~\bibnamefont
  {Becher}}\ and\ \bibinfo {author} {\bibfnamefont {G.}~\bibnamefont {Bell}},\
  }\href@noop {} {\bibfield  {journal} {\bibinfo  {journal} {Phys. Lett.}\
  }\textbf {\bibinfo {volume} {B713}},\ \bibinfo {pages} {41} (\bibinfo {year}
  {2012})},\ \Eprint {http://arxiv.org/abs/1112.3907} {arXiv:1112.3907
  [hep-ph]} \BibitemShut {NoStop}%
\bibitem [{\citenamefont {Gehrmann}\ \emph {et~al.}(2012)\citenamefont
  {Gehrmann}, \citenamefont {Lubbert},\ and\ \citenamefont
  {Yang}}]{Gehrmann:2012ze}%
  \BibitemOpen
  \bibfield  {author} {\bibinfo {author} {\bibfnamefont {T.}~\bibnamefont
  {Gehrmann}}, \bibinfo {author} {\bibfnamefont {T.}~\bibnamefont {Lubbert}}, \
  and\ \bibinfo {author} {\bibfnamefont {L.~L.}\ \bibnamefont {Yang}},\
  }\href@noop {} {\bibfield  {journal} {\bibinfo  {journal} {Phys. Rev. Lett.}\
  }\textbf {\bibinfo {volume} {109}},\ \bibinfo {pages} {242003} (\bibinfo
  {year} {2012})},\ \Eprint {http://arxiv.org/abs/1209.0682} {arXiv:1209.0682
  [hep-ph]} \BibitemShut {NoStop}%
\bibitem [{\citenamefont {Li}\ \emph {et~al.}(2016)\citenamefont {Li},
  \citenamefont {Neill},\ and\ \citenamefont {Zhu}}]{Li:2016axz}%
  \BibitemOpen
  \bibfield  {author} {\bibinfo {author} {\bibfnamefont {Y.}~\bibnamefont
  {Li}}, \bibinfo {author} {\bibfnamefont {D.}~\bibnamefont {Neill}}, \ and\
  \bibinfo {author} {\bibfnamefont {H.~X.}\ \bibnamefont {Zhu}},\ }\href@noop
  {} {\  (\bibinfo {year} {2016})},\ \Eprint {http://arxiv.org/abs/1604.00392}
  {arXiv:1604.00392 [hep-ph]} \BibitemShut {NoStop}%
\bibitem [{\citenamefont {Echevarria}\ \emph {et~al.}(2016)\citenamefont
  {Echevarria}, \citenamefont {Scimemi},\ and\ \citenamefont
  {Vladimirov}}]{Echevarria:2016scs}%
  \BibitemOpen
  \bibfield  {author} {\bibinfo {author} {\bibfnamefont {M.~G.}\ \bibnamefont
  {Echevarria}}, \bibinfo {author} {\bibfnamefont {I.}~\bibnamefont {Scimemi}},
  \ and\ \bibinfo {author} {\bibfnamefont {A.}~\bibnamefont {Vladimirov}},\
  }\href@noop {} {\bibfield  {journal} {\bibinfo  {journal} {JHEP}\ }\textbf
  {\bibinfo {volume} {09}},\ \bibinfo {pages} {004} (\bibinfo {year} {2016})},\
  \Eprint {http://arxiv.org/abs/1604.07869} {arXiv:1604.07869 [hep-ph]}
  \BibitemShut {NoStop}%
\bibitem [{\citenamefont {Ji}\ \emph {et~al.}(2005)\citenamefont {Ji},
  \citenamefont {Ma},\ and\ \citenamefont {Yuan}}]{Ji:2004wu}%
  \BibitemOpen
  \bibfield  {author} {\bibinfo {author} {\bibfnamefont {X.-d.}\ \bibnamefont
  {Ji}}, \bibinfo {author} {\bibfnamefont {J.-p.}\ \bibnamefont {Ma}}, \ and\
  \bibinfo {author} {\bibfnamefont {F.}~\bibnamefont {Yuan}},\ }\href@noop {}
  {\bibfield  {journal} {\bibinfo  {journal} {Phys. Rev.}\ }\textbf {\bibinfo
  {volume} {D71}},\ \bibinfo {pages} {034005} (\bibinfo {year} {2005})},\
  \Eprint {http://arxiv.org/abs/hep-ph/0404183} {arXiv:hep-ph/0404183 [hep-ph]}
  \BibitemShut {NoStop}%
\bibitem [{\citenamefont {Prokudin}\ \emph {et~al.}(2015)\citenamefont
  {Prokudin}, \citenamefont {Sun},\ and\ \citenamefont
  {Yuan}}]{Prokudin:2015ysa}%
  \BibitemOpen
  \bibfield  {author} {\bibinfo {author} {\bibfnamefont {A.}~\bibnamefont
  {Prokudin}}, \bibinfo {author} {\bibfnamefont {P.}~\bibnamefont {Sun}}, \
  and\ \bibinfo {author} {\bibfnamefont {F.}~\bibnamefont {Yuan}},\ }\href@noop
  {} {\bibfield  {journal} {\bibinfo  {journal} {Phys. Lett.}\ }\textbf
  {\bibinfo {volume} {B750}},\ \bibinfo {pages} {533} (\bibinfo {year}
  {2015})},\ \Eprint {http://arxiv.org/abs/1505.05588} {arXiv:1505.05588
  [hep-ph]} \BibitemShut {NoStop}%
\bibitem [{\citenamefont {Radyushkin}(2017)}]{Radyushkin:2017cyf}%
  \BibitemOpen
  \bibfield  {author} {\bibinfo {author} {\bibfnamefont {A.~V.}\ \bibnamefont
  {Radyushkin}},\ }\href@noop {} {\bibfield  {journal} {\bibinfo  {journal}
  {Phys. Rev.}\ }\textbf {\bibinfo {volume} {D96}},\ \bibinfo {pages} {034025}
  (\bibinfo {year} {2017})},\ \Eprint {http://arxiv.org/abs/1705.01488}
  {arXiv:1705.01488 [hep-ph]} \BibitemShut {NoStop}%
\bibitem [{\citenamefont {Collins}\ and\ \citenamefont
  {Tkachov}(1992)}]{Collins:1992tv}%
  \BibitemOpen
  \bibfield  {author} {\bibinfo {author} {\bibfnamefont {J.~C.}\ \bibnamefont
  {Collins}}\ and\ \bibinfo {author} {\bibfnamefont {F.~V.}\ \bibnamefont
  {Tkachov}},\ }\href@noop {} {\bibfield  {journal} {\bibinfo  {journal} {Phys.
  Lett.}\ }\textbf {\bibinfo {volume} {B294}},\ \bibinfo {pages} {403}
  (\bibinfo {year} {1992})},\ \Eprint {http://arxiv.org/abs/hep-ph/9208209}
  {arXiv:hep-ph/9208209 [hep-ph]} \BibitemShut {NoStop}%
\bibitem [{\citenamefont {Collins}(2008)}]{Collins:2008ht}%
  \BibitemOpen
  \bibfield  {author} {\bibinfo {author} {\bibfnamefont {J.}~\bibnamefont
  {Collins}},\ }\bibfield  {booktitle} {\emph {\bibinfo {booktitle}
  {{Proceedings, International Workshop on Relativistic nuclear and particle
  physics (Light Cone 2008): Mulhouse, France, July 7-11, 2008}}},\ }\href@noop
  {} {\bibfield  {journal} {\bibinfo  {journal} {PoS}\ }\textbf {\bibinfo
  {volume} {LC2008}},\ \bibinfo {pages} {028} (\bibinfo {year} {2008})},\
  \Eprint {http://arxiv.org/abs/0808.2665} {arXiv:0808.2665 [hep-ph]}
  \BibitemShut {NoStop}%
\bibitem [{\citenamefont {Chiu}\ \emph
  {et~al.}(2012{\natexlab{b}})\citenamefont {Chiu}, \citenamefont {Jain},
  \citenamefont {Neill},\ and\ \citenamefont {Rothstein}}]{Chiu:2011qc}%
  \BibitemOpen
  \bibfield  {author} {\bibinfo {author} {\bibfnamefont {J.-y.}\ \bibnamefont
  {Chiu}}, \bibinfo {author} {\bibfnamefont {A.}~\bibnamefont {Jain}}, \bibinfo
  {author} {\bibfnamefont {D.}~\bibnamefont {Neill}}, \ and\ \bibinfo {author}
  {\bibfnamefont {I.~Z.}\ \bibnamefont {Rothstein}},\ }\href@noop {} {\bibfield
   {journal} {\bibinfo  {journal} {Phys. Rev. Lett.}\ }\textbf {\bibinfo
  {volume} {108}},\ \bibinfo {pages} {151601} (\bibinfo {year}
  {2012}{\natexlab{b}})},\ \Eprint {http://arxiv.org/abs/1104.0881}
  {arXiv:1104.0881 [hep-ph]} \BibitemShut {NoStop}%
\bibitem [{\citenamefont {Beneke}\ and\ \citenamefont
  {Feldmann}(2004)}]{Beneke:2003pa}%
  \BibitemOpen
  \bibfield  {author} {\bibinfo {author} {\bibfnamefont {M.}~\bibnamefont
  {Beneke}}\ and\ \bibinfo {author} {\bibfnamefont {T.}~\bibnamefont
  {Feldmann}},\ }\href@noop {} {\bibfield  {journal} {\bibinfo  {journal}
  {Nucl. Phys.}\ }\textbf {\bibinfo {volume} {B685}},\ \bibinfo {pages} {249}
  (\bibinfo {year} {2004})},\ \Eprint {http://arxiv.org/abs/hep-ph/0311335}
  {arXiv:hep-ph/0311335 [hep-ph]} \BibitemShut {NoStop}%
\bibitem [{\citenamefont {Chiu}\ \emph {et~al.}(2008)\citenamefont {Chiu},
  \citenamefont {Golf}, \citenamefont {Kelley},\ and\ \citenamefont
  {Manohar}}]{Chiu:2007yn}%
  \BibitemOpen
  \bibfield  {author} {\bibinfo {author} {\bibfnamefont {J.-y.}\ \bibnamefont
  {Chiu}}, \bibinfo {author} {\bibfnamefont {F.}~\bibnamefont {Golf}}, \bibinfo
  {author} {\bibfnamefont {R.}~\bibnamefont {Kelley}}, \ and\ \bibinfo {author}
  {\bibfnamefont {A.~V.}\ \bibnamefont {Manohar}},\ }\href@noop {} {\bibfield
  {journal} {\bibinfo  {journal} {Phys. Rev. Lett.}\ }\textbf {\bibinfo
  {volume} {100}},\ \bibinfo {pages} {021802} (\bibinfo {year} {2008})},\
  \Eprint {http://arxiv.org/abs/0709.2377} {arXiv:0709.2377 [hep-ph]}
  \BibitemShut {NoStop}%
\bibitem [{\citenamefont {Chiu}\ \emph {et~al.}(2009)\citenamefont {Chiu},
  \citenamefont {Fuhrer}, \citenamefont {Hoang}, \citenamefont {Kelley},\ and\
  \citenamefont {Manohar}}]{Chiu:2009yx}%
  \BibitemOpen
  \bibfield  {author} {\bibinfo {author} {\bibfnamefont {J.-y.}\ \bibnamefont
  {Chiu}}, \bibinfo {author} {\bibfnamefont {A.}~\bibnamefont {Fuhrer}},
  \bibinfo {author} {\bibfnamefont {A.~H.}\ \bibnamefont {Hoang}}, \bibinfo
  {author} {\bibfnamefont {R.}~\bibnamefont {Kelley}}, \ and\ \bibinfo {author}
  {\bibfnamefont {A.~V.}\ \bibnamefont {Manohar}},\ }\href@noop {} {\bibfield
  {journal} {\bibinfo  {journal} {Phys. Rev.}\ }\textbf {\bibinfo {volume}
  {D79}},\ \bibinfo {pages} {053007} (\bibinfo {year} {2009})},\ \Eprint
  {http://arxiv.org/abs/0901.1332} {arXiv:0901.1332 [hep-ph]} \BibitemShut
  {NoStop}%
\bibitem [{\citenamefont {Stewart}\ \emph {et~al.}(2010)\citenamefont
  {Stewart}, \citenamefont {Tackmann},\ and\ \citenamefont
  {Waalewijn}}]{Stewart:2009yx}%
  \BibitemOpen
  \bibfield  {author} {\bibinfo {author} {\bibfnamefont {I.~W.}\ \bibnamefont
  {Stewart}}, \bibinfo {author} {\bibfnamefont {F.~J.}\ \bibnamefont
  {Tackmann}}, \ and\ \bibinfo {author} {\bibfnamefont {W.~J.}\ \bibnamefont
  {Waalewijn}},\ }\href@noop {} {\bibfield  {journal} {\bibinfo  {journal}
  {Phys. Rev. D}\ }\textbf {\bibinfo {volume} {81}},\ \bibinfo {pages} {094035}
  (\bibinfo {year} {2010})},\ \Eprint {http://arxiv.org/abs/0910.0467}
  {arXiv:0910.0467} \BibitemShut {NoStop}%
\bibitem [{\citenamefont {Buffing}\ \emph {et~al.}(2018)\citenamefont
  {Buffing}, \citenamefont {Diehl},\ and\ \citenamefont
  {Kasemets}}]{Buffing:2017mqm}%
  \BibitemOpen
  \bibfield  {author} {\bibinfo {author} {\bibfnamefont {M.~G.~A.}\
  \bibnamefont {Buffing}}, \bibinfo {author} {\bibfnamefont {M.}~\bibnamefont
  {Diehl}}, \ and\ \bibinfo {author} {\bibfnamefont {T.}~\bibnamefont
  {Kasemets}},\ }\href@noop {} {\bibfield  {journal} {\bibinfo  {journal}
  {JHEP}\ }\textbf {\bibinfo {volume} {01}},\ \bibinfo {pages} {044} (\bibinfo
  {year} {2018})},\ \Eprint {http://arxiv.org/abs/1708.03528} {arXiv:1708.03528
  [hep-ph]} \BibitemShut {NoStop}%
\bibitem [{\citenamefont {Manohar}\ and\ \citenamefont
  {Stewart}(2007)}]{Manohar:2006nz}%
  \BibitemOpen
  \bibfield  {author} {\bibinfo {author} {\bibfnamefont {A.~V.}\ \bibnamefont
  {Manohar}}\ and\ \bibinfo {author} {\bibfnamefont {I.~W.}\ \bibnamefont
  {Stewart}},\ }\href@noop {} {\bibfield  {journal} {\bibinfo  {journal} {Phys.
  Rev.}\ }\textbf {\bibinfo {volume} {D76}},\ \bibinfo {pages} {074002}
  (\bibinfo {year} {2007})},\ \Eprint {http://arxiv.org/abs/hep-ph/0605001}
  {arXiv:hep-ph/0605001 [hep-ph]} \BibitemShut {NoStop}%
\bibitem [{\citenamefont {Li}\ and\ \citenamefont {Zhu}(2017)}]{Li:2016ctv}%
  \BibitemOpen
  \bibfield  {author} {\bibinfo {author} {\bibfnamefont {Y.}~\bibnamefont
  {Li}}\ and\ \bibinfo {author} {\bibfnamefont {H.~X.}\ \bibnamefont {Zhu}},\
  }\href@noop {} {\bibfield  {journal} {\bibinfo  {journal} {Phys. Rev. Lett.}\
  }\textbf {\bibinfo {volume} {118}},\ \bibinfo {pages} {022004} (\bibinfo
  {year} {2017})},\ \Eprint {http://arxiv.org/abs/1604.01404} {arXiv:1604.01404
  [hep-ph]} \BibitemShut {NoStop}%
\bibitem [{\citenamefont {Ebert}\ and\ \citenamefont
  {Tackmann}(2017)}]{Ebert:2016gcn}%
  \BibitemOpen
  \bibfield  {author} {\bibinfo {author} {\bibfnamefont {M.~A.}\ \bibnamefont
  {Ebert}}\ and\ \bibinfo {author} {\bibfnamefont {F.~J.}\ \bibnamefont
  {Tackmann}},\ }\href@noop {} {\bibfield  {journal} {\bibinfo  {journal}
  {JHEP}\ }\textbf {\bibinfo {volume} {02}},\ \bibinfo {pages} {110} (\bibinfo
  {year} {2017})},\ \Eprint {http://arxiv.org/abs/1611.08610} {arXiv:1611.08610
  [hep-ph]} \BibitemShut {NoStop}%
\bibitem [{\citenamefont {Scimemi}\ and\ \citenamefont
  {Vladimirov}(2018{\natexlab{b}})}]{Scimemi:2018xaf}%
  \BibitemOpen
  \bibfield  {author} {\bibinfo {author} {\bibfnamefont {I.}~\bibnamefont
  {Scimemi}}\ and\ \bibinfo {author} {\bibfnamefont {A.}~\bibnamefont
  {Vladimirov}},\ }\href@noop {} {\bibfield  {journal} {\bibinfo  {journal}
  {JHEP}\ }\textbf {\bibinfo {volume} {08}},\ \bibinfo {pages} {003} (\bibinfo
  {year} {2018}{\natexlab{b}})},\ \Eprint {http://arxiv.org/abs/1803.11089}
  {arXiv:1803.11089 [hep-ph]} \BibitemShut {NoStop}%
\bibitem [{\citenamefont {Pietrulewicz}\ \emph {et~al.}(2017)\citenamefont
  {Pietrulewicz}, \citenamefont {Samitz}, \citenamefont {Spiering},\ and\
  \citenamefont {Tackmann}}]{Pietrulewicz:2017gxc}%
  \BibitemOpen
  \bibfield  {author} {\bibinfo {author} {\bibfnamefont {P.}~\bibnamefont
  {Pietrulewicz}}, \bibinfo {author} {\bibfnamefont {D.}~\bibnamefont
  {Samitz}}, \bibinfo {author} {\bibfnamefont {A.}~\bibnamefont {Spiering}}, \
  and\ \bibinfo {author} {\bibfnamefont {F.~J.}\ \bibnamefont {Tackmann}},\
  }\href@noop {} {\bibfield  {journal} {\bibinfo  {journal} {JHEP}\ }\textbf
  {\bibinfo {volume} {08}},\ \bibinfo {pages} {114} (\bibinfo {year} {2017})},\
  \Eprint {http://arxiv.org/abs/1703.09702} {arXiv:1703.09702 [hep-ph]}
  \BibitemShut {NoStop}%
\bibitem [{\citenamefont {Vladimirov}(2017)}]{Vladimirov:2016dll}%
  \BibitemOpen
  \bibfield  {author} {\bibinfo {author} {\bibfnamefont {A.~A.}\ \bibnamefont
  {Vladimirov}},\ }\href@noop {} {\bibfield  {journal} {\bibinfo  {journal}
  {Phys. Rev. Lett.}\ }\textbf {\bibinfo {volume} {118}},\ \bibinfo {pages}
  {062001} (\bibinfo {year} {2017})},\ \Eprint
  {http://arxiv.org/abs/1610.05791} {arXiv:1610.05791 [hep-ph]} \BibitemShut
  {NoStop}%
\bibitem [{\citenamefont {Ma}\ and\ \citenamefont
  {Qiu}(2018{\natexlab{a}})}]{Ma:2014jla}%
  \BibitemOpen
  \bibfield  {author} {\bibinfo {author} {\bibfnamefont {Y.-Q.}\ \bibnamefont
  {Ma}}\ and\ \bibinfo {author} {\bibfnamefont {J.-W.}\ \bibnamefont {Qiu}},\
  }\href@noop {} {\bibfield  {journal} {\bibinfo  {journal} {Phys. Rev.}\
  }\textbf {\bibinfo {volume} {D98}},\ \bibinfo {pages} {074021} (\bibinfo
  {year} {2018}{\natexlab{a}})},\ \Eprint {http://arxiv.org/abs/1404.6860}
  {arXiv:1404.6860 [hep-ph]} \BibitemShut {NoStop}%
\bibitem [{\citenamefont {Ma}\ and\ \citenamefont
  {Qiu}(2018{\natexlab{b}})}]{Ma:2017pxb}%
  \BibitemOpen
  \bibfield  {author} {\bibinfo {author} {\bibfnamefont {Y.-Q.}\ \bibnamefont
  {Ma}}\ and\ \bibinfo {author} {\bibfnamefont {J.-W.}\ \bibnamefont {Qiu}},\
  }\href@noop {} {\bibfield  {journal} {\bibinfo  {journal} {Phys. Rev. Lett.}\
  }\textbf {\bibinfo {volume} {120}},\ \bibinfo {pages} {022003} (\bibinfo
  {year} {2018}{\natexlab{b}})},\ \Eprint {http://arxiv.org/abs/1709.03018}
  {arXiv:1709.03018 [hep-ph]} \BibitemShut {NoStop}%
\bibitem [{\citenamefont {Izubuchi}\ \emph {et~al.}(2018)\citenamefont
  {Izubuchi}, \citenamefont {Ji}, \citenamefont {Jin}, \citenamefont
  {Stewart},\ and\ \citenamefont {Zhao}}]{Izubuchi:2018srq}%
  \BibitemOpen
  \bibfield  {author} {\bibinfo {author} {\bibfnamefont {T.}~\bibnamefont
  {Izubuchi}}, \bibinfo {author} {\bibfnamefont {X.}~\bibnamefont {Ji}},
  \bibinfo {author} {\bibfnamefont {L.}~\bibnamefont {Jin}}, \bibinfo {author}
  {\bibfnamefont {I.~W.}\ \bibnamefont {Stewart}}, \ and\ \bibinfo {author}
  {\bibfnamefont {Y.}~\bibnamefont {Zhao}},\ }\href@noop {} {\bibfield
  {journal} {\bibinfo  {journal} {Phys. Rev.}\ }\textbf {\bibinfo {volume}
  {D98}},\ \bibinfo {pages} {056004} (\bibinfo {year} {2018})},\ \Eprint
  {http://arxiv.org/abs/1801.03917} {arXiv:1801.03917 [hep-ph]} \BibitemShut
  {NoStop}%
\bibitem [{\citenamefont {Constantinou}\ and\ \citenamefont
  {Panagopoulos}(2017)}]{Constantinou:2017sej}%
  \BibitemOpen
  \bibfield  {author} {\bibinfo {author} {\bibfnamefont {M.}~\bibnamefont
  {Constantinou}}\ and\ \bibinfo {author} {\bibfnamefont {H.}~\bibnamefont
  {Panagopoulos}},\ }\href@noop {} {\bibfield  {journal} {\bibinfo  {journal}
  {Phys. Rev.}\ }\textbf {\bibinfo {volume} {D96}},\ \bibinfo {pages} {054506}
  (\bibinfo {year} {2017})},\ \Eprint {http://arxiv.org/abs/1705.11193}
  {arXiv:1705.11193 [hep-lat]} \BibitemShut {NoStop}%
\bibitem [{\citenamefont {Green}\ \emph {et~al.}(2018)\citenamefont {Green},
  \citenamefont {Jansen},\ and\ \citenamefont {Steffens}}]{Green:2017xeu}%
  \BibitemOpen
  \bibfield  {author} {\bibinfo {author} {\bibfnamefont {J.}~\bibnamefont
  {Green}}, \bibinfo {author} {\bibfnamefont {K.}~\bibnamefont {Jansen}}, \
  and\ \bibinfo {author} {\bibfnamefont {F.}~\bibnamefont {Steffens}},\
  }\href@noop {} {\bibfield  {journal} {\bibinfo  {journal} {Phys. Rev. Lett.}\
  }\textbf {\bibinfo {volume} {121}},\ \bibinfo {pages} {022004} (\bibinfo
  {year} {2018})},\ \Eprint {http://arxiv.org/abs/1707.07152} {arXiv:1707.07152
  [hep-lat]} \BibitemShut {NoStop}%
\bibitem [{\citenamefont {Chen}\ \emph {et~al.}(2017)\citenamefont {Chen},
  \citenamefont {Ishikawa}, \citenamefont {Jin}, \citenamefont {Lin},
  \citenamefont {Yang}, \citenamefont {Zhang},\ and\ \citenamefont
  {Zhao}}]{Chen:2017mie}%
  \BibitemOpen
  \bibfield  {author} {\bibinfo {author} {\bibfnamefont {J.-W.}\ \bibnamefont
  {Chen}}, \bibinfo {author} {\bibfnamefont {T.}~\bibnamefont {Ishikawa}},
  \bibinfo {author} {\bibfnamefont {L.}~\bibnamefont {Jin}}, \bibinfo {author}
  {\bibfnamefont {H.-W.}\ \bibnamefont {Lin}}, \bibinfo {author} {\bibfnamefont
  {Y.-B.}\ \bibnamefont {Yang}}, \bibinfo {author} {\bibfnamefont {J.-H.}\
  \bibnamefont {Zhang}}, \ and\ \bibinfo {author} {\bibfnamefont
  {Y.}~\bibnamefont {Zhao}},\ }\href@noop {} {\  (\bibinfo {year} {2017})},\
  \Eprint {http://arxiv.org/abs/1710.01089} {arXiv:1710.01089 [hep-lat]}
  \BibitemShut {NoStop}%
\bibitem [{\citenamefont {Ji}\ and\ \citenamefont {Yuan}(2002)}]{Ji:2002aa}%
  \BibitemOpen
  \bibfield  {author} {\bibinfo {author} {\bibfnamefont {X.-d.}\ \bibnamefont
  {Ji}}\ and\ \bibinfo {author} {\bibfnamefont {F.}~\bibnamefont {Yuan}},\
  }\href@noop {} {\bibfield  {journal} {\bibinfo  {journal} {Phys. Lett.}\
  }\textbf {\bibinfo {volume} {B543}},\ \bibinfo {pages} {66} (\bibinfo {year}
  {2002})},\ \Eprint {http://arxiv.org/abs/hep-ph/0206057}
  {arXiv:hep-ph/0206057 [hep-ph]} \BibitemShut {NoStop}%
\bibitem [{\citenamefont {Belitsky}\ \emph {et~al.}(2003)\citenamefont
  {Belitsky}, \citenamefont {Ji},\ and\ \citenamefont
  {Yuan}}]{Belitsky:2002sm}%
  \BibitemOpen
  \bibfield  {author} {\bibinfo {author} {\bibfnamefont {A.~V.}\ \bibnamefont
  {Belitsky}}, \bibinfo {author} {\bibfnamefont {X.}~\bibnamefont {Ji}}, \ and\
  \bibinfo {author} {\bibfnamefont {F.}~\bibnamefont {Yuan}},\ }\href@noop {}
  {\bibfield  {journal} {\bibinfo  {journal} {Nucl. Phys.}\ }\textbf {\bibinfo
  {volume} {B656}},\ \bibinfo {pages} {165} (\bibinfo {year} {2003})},\ \Eprint
  {http://arxiv.org/abs/hep-ph/0208038} {arXiv:hep-ph/0208038 [hep-ph]}
  \BibitemShut {NoStop}%
\bibitem [{\citenamefont {Idilbi}\ and\ \citenamefont
  {Scimemi}(2011)}]{Idilbi:2010im}%
  \BibitemOpen
  \bibfield  {author} {\bibinfo {author} {\bibfnamefont {A.}~\bibnamefont
  {Idilbi}}\ and\ \bibinfo {author} {\bibfnamefont {I.}~\bibnamefont
  {Scimemi}},\ }\href@noop {} {\bibfield  {journal} {\bibinfo  {journal} {Phys.
  Lett.}\ }\textbf {\bibinfo {volume} {B695}},\ \bibinfo {pages} {463}
  (\bibinfo {year} {2011})},\ \Eprint {http://arxiv.org/abs/1009.2776}
  {arXiv:1009.2776 [hep-ph]} \BibitemShut {NoStop}%
\bibitem [{\citenamefont {Garcia-Echevarria}\ \emph {et~al.}(2011)\citenamefont
  {Garcia-Echevarria}, \citenamefont {Idilbi},\ and\ \citenamefont
  {Scimemi}}]{GarciaEchevarria:2011md}%
  \BibitemOpen
  \bibfield  {author} {\bibinfo {author} {\bibfnamefont {M.}~\bibnamefont
  {Garcia-Echevarria}}, \bibinfo {author} {\bibfnamefont {A.}~\bibnamefont
  {Idilbi}}, \ and\ \bibinfo {author} {\bibfnamefont {I.}~\bibnamefont
  {Scimemi}},\ }\href@noop {} {\bibfield  {journal} {\bibinfo  {journal} {Phys.
  Rev.}\ }\textbf {\bibinfo {volume} {D84}},\ \bibinfo {pages} {011502}
  (\bibinfo {year} {2011})},\ \Eprint {http://arxiv.org/abs/1104.0686}
  {arXiv:1104.0686 [hep-ph]} \BibitemShut {NoStop}%
\bibitem [{\citenamefont {Stewart}\ and\ \citenamefont
  {Zhao}(2018)}]{Stewart:2017tvs}%
  \BibitemOpen
  \bibfield  {author} {\bibinfo {author} {\bibfnamefont {I.~W.}\ \bibnamefont
  {Stewart}}\ and\ \bibinfo {author} {\bibfnamefont {Y.}~\bibnamefont {Zhao}},\
  }\href@noop {} {\bibfield  {journal} {\bibinfo  {journal} {Phys. Rev.}\
  }\textbf {\bibinfo {volume} {D97}},\ \bibinfo {pages} {054512} (\bibinfo
  {year} {2018})},\ \Eprint {http://arxiv.org/abs/1709.04933} {arXiv:1709.04933
  [hep-ph]} \BibitemShut {NoStop}%
\bibitem [{\citenamefont {Monahan}\ and\ \citenamefont
  {Orginos}(2017)}]{Monahan:2016bvm}%
  \BibitemOpen
  \bibfield  {author} {\bibinfo {author} {\bibfnamefont {C.}~\bibnamefont
  {Monahan}}\ and\ \bibinfo {author} {\bibfnamefont {K.}~\bibnamefont
  {Orginos}},\ }\href {\doibase 10.1007/JHEP03(2017)116} {\bibfield  {journal}
  {\bibinfo  {journal} {JHEP}\ }\textbf {\bibinfo {volume} {03}},\ \bibinfo
  {pages} {116} (\bibinfo {year} {2017})},\ \Eprint
  {http://arxiv.org/abs/1612.01584} {arXiv:1612.01584 [hep-lat]} \BibitemShut
  {NoStop}%
\bibitem [{\citenamefont {Ji}\ \emph {et~al.}(2018{\natexlab{b}})\citenamefont
  {Ji}, \citenamefont {Zhang},\ and\ \citenamefont {Zhao}}]{Ji:2017oey}%
  \BibitemOpen
  \bibfield  {author} {\bibinfo {author} {\bibfnamefont {X.}~\bibnamefont
  {Ji}}, \bibinfo {author} {\bibfnamefont {J.-H.}\ \bibnamefont {Zhang}}, \
  and\ \bibinfo {author} {\bibfnamefont {Y.}~\bibnamefont {Zhao}},\ }\href@noop
  {} {\bibfield  {journal} {\bibinfo  {journal} {Phys. Rev. Lett.}\ }\textbf
  {\bibinfo {volume} {120}},\ \bibinfo {pages} {112001} (\bibinfo {year}
  {2018}{\natexlab{b}})},\ \Eprint {http://arxiv.org/abs/1706.08962}
  {arXiv:1706.08962 [hep-ph]} \BibitemShut {NoStop}%
\bibitem [{\citenamefont {Ishikawa}\ \emph {et~al.}(2017)\citenamefont
  {Ishikawa}, \citenamefont {Ma}, \citenamefont {Qiu},\ and\ \citenamefont
  {Yoshida}}]{Ishikawa:2017faj}%
  \BibitemOpen
  \bibfield  {author} {\bibinfo {author} {\bibfnamefont {T.}~\bibnamefont
  {Ishikawa}}, \bibinfo {author} {\bibfnamefont {Y.-Q.}\ \bibnamefont {Ma}},
  \bibinfo {author} {\bibfnamefont {J.-W.}\ \bibnamefont {Qiu}}, \ and\
  \bibinfo {author} {\bibfnamefont {S.}~\bibnamefont {Yoshida}},\ }\href@noop
  {} {\bibfield  {journal} {\bibinfo  {journal} {Phys. Rev.}\ }\textbf
  {\bibinfo {volume} {D96}},\ \bibinfo {pages} {094019} (\bibinfo {year}
  {2017})},\ \Eprint {http://arxiv.org/abs/1707.03107} {arXiv:1707.03107
  [hep-ph]} \BibitemShut {NoStop}%
\bibitem [{\citenamefont {Chen}\ \emph {et~al.}(2016)\citenamefont {Chen},
  \citenamefont {Cohen}, \citenamefont {Ji}, \citenamefont {Lin},\ and\
  \citenamefont {Zhang}}]{Chen:2016utp}%
  \BibitemOpen
  \bibfield  {author} {\bibinfo {author} {\bibfnamefont {J.-W.}\ \bibnamefont
  {Chen}}, \bibinfo {author} {\bibfnamefont {S.~D.}\ \bibnamefont {Cohen}},
  \bibinfo {author} {\bibfnamefont {X.}~\bibnamefont {Ji}}, \bibinfo {author}
  {\bibfnamefont {H.-W.}\ \bibnamefont {Lin}}, \ and\ \bibinfo {author}
  {\bibfnamefont {J.-H.}\ \bibnamefont {Zhang}},\ }\href@noop {} {\bibfield
  {journal} {\bibinfo  {journal} {Nucl. Phys.}\ }\textbf {\bibinfo {volume}
  {B911}},\ \bibinfo {pages} {246} (\bibinfo {year} {2016})},\ \Eprint
  {http://arxiv.org/abs/1603.06664} {arXiv:1603.06664 [hep-ph]} \BibitemShut
  {NoStop}%
\bibitem [{\citenamefont {Xiong}\ \emph {et~al.}(2014)\citenamefont {Xiong},
  \citenamefont {Ji}, \citenamefont {Zhang},\ and\ \citenamefont
  {Zhao}}]{Xiong:2013bka}%
  \BibitemOpen
  \bibfield  {author} {\bibinfo {author} {\bibfnamefont {X.}~\bibnamefont
  {Xiong}}, \bibinfo {author} {\bibfnamefont {X.}~\bibnamefont {Ji}}, \bibinfo
  {author} {\bibfnamefont {J.-H.}\ \bibnamefont {Zhang}}, \ and\ \bibinfo
  {author} {\bibfnamefont {Y.}~\bibnamefont {Zhao}},\ }\href@noop {} {\bibfield
   {journal} {\bibinfo  {journal} {Phys. Rev.}\ }\textbf {\bibinfo {volume}
  {D90}},\ \bibinfo {pages} {014051} (\bibinfo {year} {2014})},\ \Eprint
  {http://arxiv.org/abs/1310.7471} {arXiv:1310.7471 [hep-ph]} \BibitemShut
  {NoStop}%
\bibitem [{\citenamefont {Boels}\ \emph {et~al.}(2017)\citenamefont {Boels},
  \citenamefont {Huber},\ and\ \citenamefont {Yang}}]{Boels:2017skl}%
  \BibitemOpen
  \bibfield  {author} {\bibinfo {author} {\bibfnamefont {R.~H.}\ \bibnamefont
  {Boels}}, \bibinfo {author} {\bibfnamefont {T.}~\bibnamefont {Huber}}, \ and\
  \bibinfo {author} {\bibfnamefont {G.}~\bibnamefont {Yang}},\ }\href@noop {}
  {\bibfield  {journal} {\bibinfo  {journal} {Phys. Rev. Lett.}\ }\textbf
  {\bibinfo {volume} {119}},\ \bibinfo {pages} {201601} (\bibinfo {year}
  {2017})},\ \Eprint {http://arxiv.org/abs/1705.03444} {arXiv:1705.03444
  [hep-th]} \BibitemShut {NoStop}%
\bibitem [{\citenamefont {Moch}\ \emph {et~al.}(2017)\citenamefont {Moch},
  \citenamefont {Ruijl}, \citenamefont {Ueda}, \citenamefont {Vermaseren},\
  and\ \citenamefont {Vogt}}]{Moch:2017uml}%
  \BibitemOpen
  \bibfield  {author} {\bibinfo {author} {\bibfnamefont {S.}~\bibnamefont
  {Moch}}, \bibinfo {author} {\bibfnamefont {B.}~\bibnamefont {Ruijl}},
  \bibinfo {author} {\bibfnamefont {T.}~\bibnamefont {Ueda}}, \bibinfo {author}
  {\bibfnamefont {J.~A.~M.}\ \bibnamefont {Vermaseren}}, \ and\ \bibinfo
  {author} {\bibfnamefont {A.}~\bibnamefont {Vogt}},\ }\href@noop {} {\bibfield
   {journal} {\bibinfo  {journal} {JHEP}\ }\textbf {\bibinfo {volume} {10}},\
  \bibinfo {pages} {041} (\bibinfo {year} {2017})},\ \Eprint
  {http://arxiv.org/abs/1707.08315} {arXiv:1707.08315 [hep-ph]} \BibitemShut
  {NoStop}%
\bibitem [{\citenamefont {Grozin}\ \emph {et~al.}(2017)\citenamefont {Grozin},
  \citenamefont {Henn},\ and\ \citenamefont {Stahlhofen}}]{Grozin:2017css}%
  \BibitemOpen
  \bibfield  {author} {\bibinfo {author} {\bibfnamefont {A.}~\bibnamefont
  {Grozin}}, \bibinfo {author} {\bibfnamefont {J.}~\bibnamefont {Henn}}, \ and\
  \bibinfo {author} {\bibfnamefont {M.}~\bibnamefont {Stahlhofen}},\
  }\href@noop {} {\bibfield  {journal} {\bibinfo  {journal} {JHEP}\ }\textbf
  {\bibinfo {volume} {10}},\ \bibinfo {pages} {052} (\bibinfo {year} {2017})},\
  \Eprint {http://arxiv.org/abs/1708.01221} {arXiv:1708.01221 [hep-ph]}
  \BibitemShut {NoStop}%
\bibitem [{\citenamefont {Becher}\ and\ \citenamefont
  {Bell}(2014)}]{Becher:2013iya}%
  \BibitemOpen
  \bibfield  {author} {\bibinfo {author} {\bibfnamefont {T.}~\bibnamefont
  {Becher}}\ and\ \bibinfo {author} {\bibfnamefont {G.}~\bibnamefont {Bell}},\
  }\href@noop {} {\bibfield  {journal} {\bibinfo  {journal} {Phys. Rev. Lett.}\
  }\textbf {\bibinfo {volume} {112}},\ \bibinfo {pages} {182002} (\bibinfo
  {year} {2014})},\ \Eprint {http://arxiv.org/abs/1312.5327} {arXiv:1312.5327
  [hep-ph]} \BibitemShut {NoStop}%
\bibitem [{\citenamefont {Scimemi}\ and\ \citenamefont
  {Vladimirov}(2017)}]{Scimemi:2016ffw}%
  \BibitemOpen
  \bibfield  {author} {\bibinfo {author} {\bibfnamefont {I.}~\bibnamefont
  {Scimemi}}\ and\ \bibinfo {author} {\bibfnamefont {A.}~\bibnamefont
  {Vladimirov}},\ }\href@noop {} {\bibfield  {journal} {\bibinfo  {journal}
  {JHEP}\ }\textbf {\bibinfo {volume} {03}},\ \bibinfo {pages} {002} (\bibinfo
  {year} {2017})},\ \Eprint {http://arxiv.org/abs/1609.06047} {arXiv:1609.06047
  [hep-ph]} \BibitemShut {NoStop}%
\end{thebibliography}%

\end{document}